\documentclass[%
 aip,
 amsmath,amssymb,
 tightenlines, % added -- for single column
author-year,% %uncommented
]{revtex4-1}

\usepackage{graphicx}% Include figure files
\usepackage{dcolumn}% Align table columns on decimal point
\usepackage{bm}% bold math
\usepackage[utf8]{inputenc}
\usepackage[T1]{fontenc}
\usepackage{mathptmx}
\usepackage{etoolbox}

\usepackage{xcolor}
\usepackage{enumitem}
\usepackage[colorlinks=true,linkcolor=violet,citecolor=teal,urlcolor=blue]{hyperref}
\setlist{nosep} % or noitemsep

\newcommand{\raisesubfiglabel}[3]{%
\hspace{-#2}\makebox[0pt][l]{\raisebox{#1}{\bf #3}}\hspace*{#2}}

\usepackage[rightlabels]{titletoc}
\titlecontents{section}[1.4cm]{\vspace{0.mm}}%
  {{\bfseries\contentslabel{1.4cm}\hspace{0.5mm}}}%numbered sections
  {}%numberless sections -- doesn't seem to work with RevTex
  {\hspace{1.5mm}\titlerule*[2mm]{{\normalfont .}}\contentspage}

\titlecontents{subsection}[1.4cm]{\vspace{-0.5mm}}%
  {\thecontentslabel . \hspace{1mm}}%numbered subsections
  {}%numberless subsection
  {\hspace{1.5mm}\titlerule*[2mm]{.}\contentspage}%

\makeatletter
\def\@email#1#2{%
 \endgroup
 \patchcmd{\titleblock@produce}
  {\frontmatter@RRAPformat}
  {\frontmatter@RRAPformat{\produce@RRAP{*#1\href{mailto:#2}{#2}}}\frontmatter@RRAPformat}
  {}{}
}%
\makeatother
\begin{document}

\preprint{AIP/123-QED}

\title[Smoothing away grid instability and noise in PIC]{ 
  Suppressing grid instability and noise 
  in particle-in-cell simulation by smoothing}
% Force line breaks with \\
\author{Gregory R. Werner}
\affiliation{Center for Integrated Plasma Studies, Physics Department, University of Colorado Boulder, Boulder, CO 80309.}
 \email{Greg.Werner@colorado.edu}

\author{Luke C. Adams}
\affiliation{Center for Integrated Plasma Studies, Physics Department, University of Colorado Boulder, Boulder, CO 80309.}
% \homepage{http://www.Second.institution.edu/~Charlie.Author.}
%\affiliation{%
%Second institution and/or address%\\This line break forced% with \\
%}%

\author{John R. Cary}%
\affiliation{Center for Integrated Plasma Studies, Physics Department, University of Colorado Boulder, Boulder, CO 80309.}
% \altaffiliation[Also at ]{Tech-X Corp., 5621 Arapahoe Ave, Boulder, CO 80303.}%Lines break automatically or can be forced with \\
\affiliation{Tech-X Corp., 5621 Arapahoe Ave, Boulder, CO 80303.%Lines break automatically or can be forced with \\
}%

\date{\today}% It is always \today, today,

\begin{abstract}
Smoothing short-wavelength charge density variations can stabilize
explicit electrostatic particle-in-cell (PIC) plasma simulations 
against grid heating and cold beam instabilities, 
which cause unphysical heating when the Debye length is poorly resolved. 
We demonstrate this by solving the dispersion and by running 1D 
electrostatic PIC 
simulations, using an efficient smoothing algorithm that
leverages the Poisson solve.
To ensure stability, the smoothing radius must increase
with the number of Debye lengths per cell.
Smoothing also suppresses particle noise, which is 
severely exacerbated by poor resolution of the Debye length.
To help determine optimal PIC configuration,
we empirically characterize electric field noise, particle velocity 
diffusion, and unphysical energy exchanges in 1D PIC simulation, 
as a function of Debye-length resolution, smoothing, and particles per 
cell.
We also show how PIC noise causes test particles to exhibit misleading behavior.
Since smoothing reduces the effective resolution, the optimal cell size
is less than the desired resolution but can be much greater than
the Debye length, reducing computational expense.

\end{abstract}

%\maketitle
% N.B. Original template just has above: I added following hack to start toc on abstract page
{
\let\clearpage\relax
\maketitle
}
\vspace{-0.5in}

\tableofcontents

\section{\label{sec:intro} Introduction}

Particle-in-cell (PIC) simulation has become an invaluable
tool for
studying plasma behavior across a wide range of fields, from 
beam physics to astrophysics.  Its main advantage, the ability to
simulate classical plasma physics from first principles, is closely
related to one of its main frustrations: it can be unstable when 
plasma microscales (such as the electron Debye length $\lambda_D$ or 
plasma frequency $\omega_p$) are not 
resolved, even when the associated phenomena (e.g., Debye shielding, Langmuir waves) are
expected to be absent in a particular application.
Furthermore, PIC can be noisy, sometimes to the point of precluding feasible simulation.
In this paper, we will quantify how under-resolving the Debye length
introduces numerical instabilities and exacerbates noise,
and we will show how both these problems can be simply and efficiently solved
in standard explicit electrostatic PIC by smoothing the charge density
(and we will calculate how much smoothing is needed).
Thus smoothing could greatly speed computation in simulations where 
Debye-scale physics is unimportant.
Although smoothing has been previously suggested as a way to 
defeat grid instability, previous works did not calculate how much 
smoothing is needed.

By ``smoothing,'' we mean filtering out short-wavelength 
variations in the charge density field.
Its effect is similar to
using larger or higher-order particle shapes \citep{Hockney-1971}, but
smoothing the field instead of individual particles is faster.

There are many variants of PIC, but the simplest
explicit electrostatic and electromagnetic algorithms (especially 
momentum-conserving PIC with 
linear weighting or first-order particle shapes)
continue to be
widely used \citep{BirdsallPIC}.  They are generally 
fastest, in terms of time steps per second, and reliably
scalable, maintaining high efficiency on massively parallel 
supercomputers.

When explicit PIC simulations fail to resolve the Debye length, 
grid instability unphysically heats the plasma
until the Debye length is resolved
\citep{Langdon-1970,Okuda-1972a,Okuda-1972b}.
Grid instabilities arise when sub-grid variations in particle density
alias to grid-resolved wavelengths, coupling different wavelengths 
that should not be
physically coupled \citep{Langdon-1970,Birdsall_Maron-1980}.
Explicit ``energy-conserving'' algorithms (which conserve energy precisely 
only in the limit of zero time step) avoid grid instability when simulating
{\it stationary} plasma \citep{Lewis-1970,Langdon-1973}, and for 
that reason have recently received renewed
attention \citep{Barnes_Chacon-2021,Powis_Kaganovich-2024,Adams_etal-2025inPrep}.
However, they still suffer from cold beam instability,
a manifestation of grid instability that heats {\it drifting} plasma 
unphysically by converting bulk kinetic to thermal energy 
\citep{Okuda-1972b,Chen_etal-1974}.

Although grid instability can be disastrous for
$\Delta x \gg \lambda_D$, particle noise may
be the more pressing reason for using
a cell size $\Delta x \lesssim \lambda_D$.
For stationary plasmas, 
previous works have claimed a grid instability threshold around
$\lambda_D/\Delta x \approx 1/\pi$ or~$0.3$ 
\citep{Birdsall_Maron-1980,BirdsallPIC}, but recent studies
suggest the threshold may be closer to $\lambda_D/\Delta x \approx 0.15$
\citep{Adams_etal-2025inPrep}.
For $0.15 \lesssim \lambda_D/\Delta x \lesssim 1$,
grid instability may be hard to detect but
noise-driven (stochastic) heating can be so strong that 
it could be mistaken for grid instability.
In principle, noise-driven heating can always be reduced by using more simulated macroparticles, but (we will show) for $\Delta x \gtrsim \lambda_D$, reducing $\Delta x$ is likely a more efficient way to reduce noise.
For example, a 1D simulation with $\Delta x = \lambda_D$ and 50 macroparticles per cell has the same noise level as one
with $\Delta x = 2\lambda_D$ and 400 macroparticles per cell.
Regardless of grid instability, 
$\Delta x \lesssim \lambda_D$ 
may be desirable because it more efficiently suppresses PIC
noise.

Efforts to facilitate under-resolution of the
Debye length
have focused primarily on suppressing grid instability.
Implicit PIC methods, with true energy conservation (for finite time step)
can avoid grid instability 
\citep{Chen_etal-2011,Markidis-2011,Lapenta_Markidis-2011,Chen_Chacon-2014}.
However, energy-conserving 
implicit methods are significantly more complicated,
difficult to implement,
more time-consuming per particle-step, and less well understood;
explicit PIC is still much more widely used.
Thus we continue to strive for explicit PIC algorithms with improved
stability, especially-implemented algorithms.

Some notable explicit schemes that have been
at least partially successful in suppressing 
grid instability in PIC \citep{Birdsall_Maron-1980} include 
displacing the grid to disrupt short-wavelength modes that
have the highest unphysical growth rates
\citep[][]{Chen_etal-1974,Brackbill_Lapenta-1994},
explicit energy-conserving PIC (cited above),
high-order shape functions \citep{Fiuza_etal-2011,Shalaby_etal-2017},
reduced-order PIC that uses multiple 1D subdivisions of the primary 
3D grid \citep{Reza_etal-2022,Reza_etal-2023}, 
the Particle-In-Fourier method \citep{Evstatiev_Shadwick-2013,Mitchell_etal-2019},
and a machine-learning algorithm that takes advantage 
of phase-space information \citep{Aguilar_Markidis-2021arxiv}.

Sufficient smoothing stabilizes explicit PIC,
potentially offering simplicity, ease of implementation,
generalizability, and performance beyond any of those 
methods.
Suppressing grid instability with smoothing is not a new idea
\citep{Birdsall_Maron-1980}, although its effectiveness 
has not to our knowledge been previously demonstrated
systematically or beyond incremental improvements in stability.
For a particular smoothing kernel, we will calculate the 
smoothing radius needed to eliminate grid instability,
as a function of Debye-length resolution.
We will then characterize how smoothing affects grid 
instability growth as a function of $\lambda_D/\Delta x$ and
drift velocity for uniform plasma, for standard 1D momentum-conserving
PIC.
Perhaps more important, we characterize electric field noise
as a function of 
$\lambda_D/\Delta x$, the number of macroparticles per cell, and
the smoothing strength, and
further characterize
the velocity-space diffusion caused by PIC noise,
which results in an unphysical energy that grows linearly with time.

We will focus on the most common explicit,
non-relativistic, momentum-conserving electrostatic PIC algorithm with
linear weighting (cloud-in-cell), in 1D, 
applied to electron plasma with uniform immobile ion 
background---we will refer to this as MC-PIC 
when we wish to clarify that
results apply to this specific algorithm.
However, some results, and the general principle of smoothing,
will likely apply to other explicit 
electrostatic PIC 
algorithms, and may well extend to electromagnetic PIC algorithms.
In addition, smoothing may be useful for implicit algorithms that need
to suppress noise when under-resolving the Debye length.

Following this introduction, \S\ref{sec:noiseTheory} will present 
a simple argument to motivate
how under-resolving the Debye length can exacerbate PIC noise.
For example, it will show that (for a large class of electrostatic
PIC algorithms without smoothing)
initializing particles with random
positions will result in an unphysical (noise) electric field energy 
that exceeds the plasma energy when $\lambda_D \ll \Delta x$.

We will then turn to grid instability.
In~\S\ref{sec:dispersion}, we examine the electron-only electrostatic 
dispersion for finite~$\Delta x$ and 
estimate that, to eliminate grid instability with the proposed
smoothing kernel,
the charge density must be smoothed over a number of cells
that scales as
$\sim \Delta x/\lambda_D$.  

In~\S\ref{sec:poissonSmoothing} we describe the smoothing used in
this paper, which involves a modified Poisson solve.
This method is efficient for large kernels (needed for
$\lambda_D \ll \Delta x$), and 
is especially attractive because of its
similarity to the electrostatic Poisson solve that must already be
implemented in any electrostatic PIC code.

In~\S\ref{sec:dispersionResults} we calculate grid instability growth
rates by numerically solving the finite-$\Delta x$ dispersion 
for a Cauchy-squared velocity
distribution,
first without and then with smoothing.

We then report on noise, grid instability, and smoothing in 
actual 1D MC-PIC simulations in~\S\ref{sec:PICresults}.
We distinguish grid instability and 
noise-driven heating,
and characterize the unphysical electric field fluctuations and
the subsequent diffusion of particles in velocity space.
We demonstrate that
smoothing suppresses both grid heating and noise-driven heating,
and warn against 
``noisy starts'' and using test particles to study particle trajectories.

Having shown how to stabilize PIC simulations for a given
$\Delta x > \lambda_D$ by smoothing, we
discuss how to apply this method practically, calculating the
optimal cell size~$\Delta x$ when accounting for resolution degradation 
due to smoothing, and estimating potential speed-ups 
(\S\ref{sec:discussion}). 

Finally, we comprehensively summarize our results in~\S\ref{sec:summary} and suggest 
directions for future work.

\section{Under-resolving the Debye length exacerbates PIC noise}
\label{sec:noiseTheory}

PIC simulations suffer from ``particle noise'' due to statistical
fluctuations of the macroparticles representing the particle distribution.
When $\lambda_D \lesssim \Delta x$, this noise can become 
catastrophically disruptive.
This has been noted in the development of so-called asymptotic-preserving
schemes for $\lambda_D /\Delta x \rightarrow 0$, which conclude that 
in this limit, the Poisson equation is too sensitive to noise, and
use alternative ways to find the electric field \citep{Degond_etal-2010}.

It is illuminating to  estimate the electric field $E_{f\Delta x}$ 
resulting from charge density fluctuations over the length scale of a 
grid cell~$\Delta x$.
With ambient plasma density~$n$, we consider density fluctuations of 
amplitude $f n$; for completely uncorrelated particles, 
$f\approx M_{\rm ppc}^{-1/2}$, where $M_{\rm ppc}$ is the number of 
macroparticles per cell.
From Gauss's law, we estimate
$\nabla \cdot E_{f\Delta x} \sim E_{f\Delta x}/\Delta x
\sim efn/\epsilon_0$.
Writing $n$ in terms of plasma frequency 
$\omega_p = \sqrt{n e^2/m\epsilon_0} = v_{\rm th}/\lambda_D$
(assuming a somewhat thermal population of particles with temperature $m v_{\rm th}^2$ and charge-to-mass ratio $\pm e/m$),
\begin{eqnarray} \label{eq:randomNoise}
  E_{f\Delta x} &\sim & 
    f \frac{\omega_p m v_{\rm th}}{e}
     \frac{\Delta x}{\lambda_D}
.\end{eqnarray}

Acting on a particle over some time $\tau$, $E_{f\Delta x}$ will
cause a relative velocity change
\begin{eqnarray}
  \frac{\Delta v}{v_{\rm th}} &\sim& 
      \frac{e E_{f\Delta x} \tau }{m v_{\rm th}} 
      \sim  f
        \frac{\Delta x}{\lambda_D} \omega_p \tau  
.\end{eqnarray}
Therefore, under-resolving $\lambda_D$ amplifies the effect of 
grid-scale density fluctuations, as gauged by how much $E_{f\Delta x}$
would change a particle's velocity (relative to $v_{\rm th}$) over
a plasma period.
Smoothing the charge density over multiple cells can reduce this 
noise.

This is far from the whole story; the fact that statistical fluctuations 
at the cell scale can generate such influential fields means that
particles become strongly correlated, and in a self-consistent PIC
simulation the above estimate is very
wrong, as we will see in~\S\ref{sec:PICresults} \citep[also see][]{Touati_etal-2022,Acciarri_etal-2024a,Acciarri_etal-2024b}.

However, the above estimation applies rigorously in some situations.
For example, if we initialize a plasma simulation with uniformly 
random particle positions,
then, initially, $f \approx M_{\rm ppc}^{-1/2}$.  
This statistical variation
in density will result in spatial electric field fluctuations 
$E_{f\Delta x}$ with energy density
\begin{eqnarray}
  \frac{\epsilon_0}{2} E_{f\Delta x}^2
  \sim M_{\rm ppc}^{-1} \frac{n m v_{\rm th}^2}{2} 
  \left( \frac{\Delta x}{\lambda_D} \right)^2
.\end{eqnarray}
For example, initializing particles with independent random positions 
in a simulation with $\lambda_D = 0.1 \Delta x$ and $M_{\rm ppc}=100$
results in an
initial unphysical electric field energy rivaling the initial
plasma thermal energy.  
Such random loading would likely invalidate the simulation, unless
$M_{\rm ppc}$ is increased or $\Delta x$ decreased.
Thus non-random particle 
positioning becomes increasingly important for
$\lambda_D \lesssim \Delta x$.
Other processes external to the fundamental PIC algorithm, 
such as injecting particles from boundaries or
combining macroparticles, may also lead to similar statistical
variation.

Fluctuations in the electric field can give (somewhat)
random kicks to particles, causing stochastic diffusion in 
velocity space.  This results in unphysical heating
(as we will show in~\S\ref{sec:PICresults}) that increases strongly
with $\Delta x/\lambda_D$.
Importantly, noise-driven heating scales as $M_{\rm ppc}^{-1}$,
whereas the grid instability growth rate becomes independent
of~$M_{\rm ppc}$ in the limit of large $M_{\rm ppc}$.

\section{Estimating the smoothing needed to defeat grid instability}
\label{sec:dispersion}

In this section, we ignore PIC noise and consider the infinite-particle
limit where grid-instability theory applies. 
We conjecture that
grid instability can be suppressed by smoothing the charge density 
$\rho(x)$ in a particular way that depends on $\lambda_D/\Delta x$,
and using the smoothed charge density $\rho_{\rm sm}(x)$ to calculate
the electric potential~$\phi$. 
The smoothing will be described in terms of Fourier
components at wavenumber $k$: 
$\tilde{\rho}_k \mapsto \tilde{\rho}_{{\rm sm},k}$.

We will show that: if $\lambda_D/\Delta x > \alpha_c$ ensures stability
for some $\alpha_c$ (which is known empirically to be of order unity),
then (for~$\lambda_D/\Delta x < \alpha_c$) 
suppressing charge density variations with
$\tilde{\rho}_{{\rm sm},k} = [\lambda_D/(\alpha_c \Delta x)]^2 \tilde{\rho}_k$
will stabilize a mode with wavenumber~$k$.
However: this over-stabilizes all modes except the most
unstable
(in particular, modes with small-enough~$k$ are already stable with
$\tilde{\rho}_{{\rm sm},k} = \tilde{\rho}_k$).

After that important analytical step (fully explicated below), 
our argument relies on more empirical grounds.
First, grid instability is absent for $\alpha_c \sim 1$.
Second, the most unstable mode usually has wavenumber
$|k| \sim k_{\rm Nyq} \equiv \pi/\Delta x$
\citep{Birdsall_Maron-1980}.
Third, for accuracy, long-wavelength behavior
must remain (nearly) unchanged.
With this in mind, we consider 
\begin{eqnarray} \label{eq:rhoSmoothing}
  \tilde{\rho}_{{\rm sm},k} &=& \frac{1}{1+k^2 r_{\rm sm}^2} 
    \tilde{\rho}_k
\end{eqnarray}
which smooths over a scale $r_{\rm sm}$: 
variations with wavelengths much longer than $r_{\rm sm}$
($k^2r_{\rm sm}^2 \ll 1$) are unchanged, while shorter wavelengths are smoothed out, reduced in amplitude by$\,\sim (k r_{\rm sm})^{-2}$.
For stability at~$k_{\rm Nyq}$ we then need
$(k_{\rm Nyq}r_{\rm sm})^{-2} \sim \lambda_D^2/(\alpha_c\Delta x)^2$, hence
\begin{eqnarray} \label{eq:stabilityCriterion}
  \frac{r_{\rm sm}}{\Delta x} \sim 
  \frac{\alpha_c}{\pi} \frac{\Delta x}{\lambda_D}
.\end{eqnarray}
Then we hope that this smoothing profile will stabilize all other
modes besides~$|k|\sim k_{\rm Nyq}$.
Thus we conjecture that PIC will be stable when
smoothing according to Eqs.~(\ref{eq:rhoSmoothing}) 
and (\ref{eq:stabilityCriterion}).

The rest of this section presents a detailed motivation of this conjecture; 
while the analytical argument is illuminating, 
ultimately 
the conjecture is still a guess to be confirmed by numerical work
in later sections.  Readers less interested in the underlying motivation
may prefer to skip to the next section.

\vspace{0.05in}

In the following, we will examine the finite-difference plasma
dispersion to show how grid instability can be suppressed
by smoothing. 
For derivation of the dispersion and an introduction to
grid instability, we refer the reader to previous works
\citep{Langdon-1970,Birdsall_Maron-1980,BirdsallPIC}; 
however, for completeness,
we derive the dispersion in Appendix~\ref{sec:dispersionDerivation}.
We account for spatial discretization, but use continuous time and
particle distribution (i.e., an infinite number of particles).

To find the frequencies and growth rates of 
electrostatic (electron-only) modes in a uniform
plasma, we solve
$D(k,\omega)=0$, where the finite difference dispersion $D$ is
\citep[cf. Appendix~\ref{sec:dispersionDerivation} of this paper, 
and Sec.~8-10 of][]{BirdsallPIC}
% \psi defined in Sec.~8-3
\begin{eqnarray} \label{eq:dispersion}
  D(k,\omega) &=& 1 - 
   \frac{\omega_p^2 \kappa(k)}{K(k)^2}
      \sum_g \frac{|\tilde{S}_{k+k_g}|^2}{k+k_g}
      \int \frac{\partial_v F_0 \; dv}{ v - \omega / (k+k_g) } 
\end{eqnarray}
Everything to the right of $\omega_p^2/K^2$ will be 
relatively unimportant for our argument, but for completeness
we briefly describe the elements in the dispersion:
\begin{itemize}[leftmargin=*]
  \item Grid nodes are $x_j=j\Delta x$ for integer $j$.
  \item $\omega$ is the complex mode frequency; with time dependence
    $e^{-i\omega t}$, growth occurs if Im$[\omega]>0$.
  \item $k \in (-\pi/\Delta x, \pi/\Delta x]$ is the (real) wavenumber.
  \item $k_g = 2\pi g/\Delta x$, for integer~$g$, 
    is the spacing between wavenumbers 
    that alias to the same grid wavenumber.  E.g.,
    $\exp[i(k+k_g) j \Delta x]=\exp(ik j \Delta x)$ 
    for integers $j$ and $g$.
    The $k_g$ are needed to describe sub-grid wavelengths of the
    particle distribution $f(x,v)$ and shape function $S(x)$.
    We sometimes write (uniquely) 
    $q=k+k_g$, where $q$ ranges over all real numbers.
  \item $\omega_p^2=n e^2/(\epsilon_0 m)$ is the plasma frequency
    squared corresponding to plasma number density $n$.
  \item $m$ is the electron mass; $-e$ is the electron charge;
    $\epsilon_0$ is the vacuum permittivity.
  \item $f_0(x,v) = n F_0(v)$ is the initial, 
    unperturbed velocity distribution; 
    $F_0(v)$ is normalized to one.
  \item $S(j\Delta x;x)=S(j\Delta x-x)$ is the so-called shape 
    function \citep[cf. App.~\ref{sec:dispersionDerivation} or][]{BirdsallPIC}.
    It describes how fields are interpolated from grid node $j\Delta x$
    to position $x$, i.e., $E(x) = \sum_j E_j S(j\Delta x-x)$, and
    how particle charge is deposited to the grid:
    a particle at $x$ with charge $q$ contributes to
      charge density
      $\rho_j = (q/V)S(j\Delta x - x)$ at node $j\Delta x$, 
      where $V$ is the volume around the node.
   \item $\tilde{S}_{k+k_g}=\tilde{S}_q$ 
     is the Fourier transform of $S(x)$.
   \item $K(k)^2$ describes the ``$\phi$-solve,'' 
     generalizing
     the usual Poisson solve (solving $-\nabla^2 \phi=\rho/\epsilon_0$
     for $\phi$) to include the effect of smoothing.
     In Fourier space, 
     $K(k)^2 \tilde{\phi}_k = \tilde{\rho}_k/\epsilon_0$. 
     The physically correct $\phi$-solve has $K(k)^2 = k^2$, 
     but with  a standard 3-point
     finite difference Laplacian (without smoothing),
     $K(k)^2 = (2/\Delta x)^2 \sin^2(k\Delta x/2)$.
   \item $i\kappa(k)$ is the Fourier representation of the
     discretized
     gradient operator used to calculate $E_j$ from
     $\phi_j$.  
     The physically correct dependence is $\kappa(k) = k$;
     with a standard centered finite difference 
     (equivalent to differencing $\phi$ at
     nodes to $E$ at edge-centers, and then interpolating $E$ from
     edge-centers to nodes), $\kappa(k) = \sin (k\Delta x) / \Delta x$.
\end{itemize}
In continuous space (with translation invariance) all Fourier modes
decouple, leaving only the $g=0$ term in Eq.~(\ref{eq:dispersion});
however, the grid breaks continuous translational symmetry,
coupling modes with
$q=k + k_g$ for the same $k$ but different integers~$g$.
Grid instability arises because sub-grid density variations 
in $f(x,v)$ (with wavenumber~$k+k_g$ for $g\neq 0$) 
alias to or interact with electric fields with grid-resolved wavenumber~$k$.
This can lead to unphysical inverse Landau damping (i.e., growth).

Our analysis here is almost independent of the
details of the dispersion and therefore applies generally to a large class 
of explicit electrostatic PIC algorithms; it relies on the fact that the
dispersion can be written
\begin{eqnarray}
  D(k,\omega) &=& 1 - \frac{\omega_p^2}{K^2}
    \times 
    \left[ { \textrm{terms independent of } n 
     \atop \textrm{and $\phi$-solve} } \right]
\end{eqnarray}
The plasma density and the $\phi$-solve are completely 
specified by $\omega_p^2$ and $K(k)^2$, respectively, 
and these appear in $D(k,\omega)$
only in the form~$\omega_p^2/K^2$.
We can exploit this to determine, for a simulation of a given plasma 
density, how we have to alter the $\phi$-solve ($K^2$) to 
obtain stability.
To do this, we consider two simulations: one with the
desired $\omega_p$ and an unknown $K^2$, and another that is 
identical 
except with $\omega_p'$ such that it is stable 
with a $\phi$-solve typically used in PIC, $K'(k)^2 \approx k^2$
[i.e., the two simulations are identical in $F_0(v)$, $\Delta x$, $S(x)$, and $\kappa(k)$, but differ in density].
If $\alpha_c$ is known such that PIC is stable
when $\lambda_D' = \alpha_c \Delta x$, hence
$\omega_p' = v_{\rm th}/\lambda_D' = v_{\rm th}/(\alpha_c \Delta x)$,
we can stabilize the first simulation at a given~$k$
if we use a $\phi$-solve such that 
$K(k)^2 = (\omega_p^2 / {\omega_p'}^2) K'(k)^2 
= (\alpha_c \Delta x/\lambda_D)^2 k^2$.
These two simulations have the same dispersion;
if one has no growing modes for a given~$k$, the other also has no
growing modes.

Altering $K(k)^2$ in this way stabilizes modes with any~$k$,
but it is overkill for all but the most unstable mode.
Unfortunately, this argument does not tell us 
the smallest $K(k)^2/k^2$ that guarantees stability at other~$k$.
However, grid
instability growth rates tend to increase as $|k|$ increases,
with the fastest growth around $|k|\sim k_{\rm Nyq}=\pi/\Delta x$
\citep[cf.~\S\ref{sec:dispersionResults}; also cf.][]{Birdsall_Maron-1980}.
This suggests choosing $K(k)^2/k^2$ to increase 
smoothly from 1 (at small~$k$) to a maximum value 
$\sim (\alpha_c \Delta x/\lambda_D)^2$ at~$k_{\rm Nyq}$.

Thus we consider the candidate
\begin{eqnarray} \label{eq:stabilizingK}
  K(k)^2 &\sim & k^2 (1 + k^2 r_{\rm sm}^2)
.\end{eqnarray}
With $r_{\rm sm}$ given by Eq.~(\ref{eq:stabilityCriterion}), 
this ensures accuracy at long wavelengths and stability at $k_{\rm Nyq}$,
and we will have to verify numerically whether modes at 
all~$k$ are stabilized.

Equation~(\ref{eq:stabilizingK}) implies that the $\phi$-solve is
given by
$k^2 (1+k^2 r_{\rm sm}^2) \tilde{\phi}_k = \tilde{\rho}_k/\epsilon_0$.
This is equivalent to Eq.~(\ref{eq:rhoSmoothing}) 
combined with
$k^2 \tilde{\phi}_k \approx \tilde{\rho}_{{\rm sm},k}/\epsilon_0$.
Thus this $\phi$-solve has a nice physical interpretation:
it yields the potential (via the usual Poisson solve) of the smoothed charge distribution
$\rho_{\rm sm}(x)$.

In summary, we estimate---and will later confirm---that smoothing the charge density $\rho$ according to
Eqs.~(\ref{eq:rhoSmoothing}) and~(\ref{eq:stabilityCriterion})
will eliminate the finite grid instability.
Importantly, $r_{\rm sm}$ must increase as the Debye length becomes more 
poorly resolved.
The obvious disadvantage of this approach is that it coarsens
the
effective resolution of the simulation from $\Delta x$ to 
$r_{\rm sm}$.
We will discuss this later in~\S\ref{sec:discussion}.

We note that the dependence 
$r_{\rm sm}/\Delta x \sim \Delta x/\lambda_D$ resulted from our choice of
smoothing in Eq.~(\ref{eq:rhoSmoothing}).  Had we instead chosen
$\tilde{\rho}_{{\rm sm},k}=\tilde{\rho}_k/(1+k^4 r_{\rm sm}^4)$, 
the necessary smoothing length (to stabilize $k_{\rm Nyq}$)
would be $r_{\rm sm}/\Delta x \sim (\Delta x/\lambda_D)^{1/2}$.
This yields superior effective resolution, but it is
less likely to stabilize modes with middling~$k$, possibly requiring
larger~$r_{\rm sm}$.
Determining the optimal smoothing is left to future work.

In the next section, we suggest an efficient way to implement the
smoothing of Eq.~(\ref{eq:rhoSmoothing}).

\section{Smoothing with a modified Poisson solve}
\label{sec:poissonSmoothing}

In this paper we propose that smoothing on a length scale 
$r_{\rm sm} \sim \Delta x^2 / \lambda_D$ 
can suppress grid instability and PIC noise
to make simulation accurate for $\Delta x \gg \lambda_D$.
For large $\Delta x/\lambda_D$, the large smoothing kernel 
spans many cells and may require nontrivial computation.

Smoothing can be accomplished by several means, and it is not this 
paper's goal to find or evaluate the best one.  We merely wish to 
point out that a suitable smoothing operation can be performed with
a modified Poisson solve.  Since electrostatic PIC simulations already
must perform a Poisson solve, we hope this will at most double the
field solve time (however, 1D is special---see the note at the end of this section).
Moreover, this method offers
flexibility to handle whatever complications 
(e.g., variable mesh spacing) are already handled by the 
electrostatic solver.

Given charge density $\rho(n\Delta x)$ at grid nodes 
(with Fourier transform $\tilde{\rho}_k$)
we wish to find the smoothed density $\rho_{\rm sm}$ 
satisfying~Eq.~(\ref{eq:rhoSmoothing})
for any chosen smoothing length $r_{\rm sm}$.
This can be achieved by solving
\begin{eqnarray}
  (-\nabla^2  + r_{\rm sm}^{-2}) \rho_{\rm sm} &=& r_{\rm sm}^{-2} \rho
.\end{eqnarray}
In a PIC code this can likely be solved in the same way as the Poisson
solve for the potential~$\phi$.  It may even be faster because the
left-hand operator is more diagonally dominant, which reflects the
fact that it has a more limited range: whereas  
solutions (Green functions) of
$-\nabla^2 \phi = \delta^3(r)$ decay in 3D as $\phi \sim 1/r$, solutions of
$(-\nabla^2+r_{\rm sm}^{-2}) \phi = \delta^3(r)$ 
decay as $\phi \sim e^{-r/r_{\rm sm}}/r$.

When solving for $\rho_{\rm sm}$ by this method, one must consider
boundary conditions.  While periodic boundary conditions pose no
extra difficulty
(and, on a regular grid, allow simple solution via Fourier transform),
other electrostatic boundary conditions may require different
smoothing boundary conditions.
With the modified Poisson solution, Neumann boundary conditions,
$\hat{\bf n}\cdot \nabla \rho_{\rm sm} = 0$,
may seem appropriate  because they
conserve charge over the volume $V$, since (here,
$\hat{\bf n}$ is the boundary surface normal)
\begin{widetext}
\begin{eqnarray}
  \int_V (\rho_{\rm sm} - \rho) dV
  &\!\!=\!\!&
  \int_V [\rho_{\rm sm} - r_{\rm sm}^2 (-\nabla^2 + r_{\rm sm}^{-2})\rho_{\rm sm}] dV
  =
  \oint_{\partial V} r_{\rm sm}^2 \nabla \rho_{\rm sm} \cdot \hat{\bf n} dA
\end{eqnarray}
\end{widetext}
where the last integral is over the boundary surface of $V$.

However, other boundary conditions (such as $\rho_{\rm sm}=0$)
may be preferable.  Real boundaries tend to affect the plasma
significantly, e.g., forming sheaths on the Debye scale.
The choice to simulate with $\Delta x \gg \lambda_D$ may well preclude
accurate boundary treatment, and compared with that neglect, 
the choice of smoothing boundary condition may be unimportant.

This consideration highlights 
another potential benefit of this method: accurate 
boundary treatment with variable mesh, with $\Delta x < \lambda_D$
near the boundary and $\Delta x \gg \lambda_D$ in the interior.
The modification of the (negative) Laplacian matrix needed to apply
smoothing
simply adds $r_{\rm sm}^{-2}$ to the diagonal 
elements; it is trivial to allow
$r_{\rm sm}=r_{\rm sm}(x)$ to vary in space without destroying important 
matrix properties like symmetry.
If $r_{\rm sm} \rightarrow 0$ at the boundary, the proper boundary condition
is $\rho_{\rm sm}=\rho$.

Varying the smoothing length $r_{\rm sm}(x)$ in space
could also be used to handle plasma with
varying density as well as simulations with varying cell size.

Since the concept of a screened (Yukawa) potential is familiar to many,
we point out that the smoothed potential 
$\phi$ is the difference
between the raw potential $\phi_r$, satisfying 
$-\nabla^2 \phi_r = \rho/\epsilon_0$
and the screened potential $\phi_{\rm scr}$,
$(-\nabla^2 + r_{\rm sm}^{-2}) \phi_{\rm scr} = \rho/\epsilon_0$:
$\phi = \phi_r - \phi_{\rm scr}$.

Last, we note that in 1D (but hopefully not in 2D or 3D), the increased
diagonal dominance of the modified Poisson solve
can hurt performance, because (at least for uniform grids)
in 1D, direct integration is likely the fastest way to solve the Poisson 
equation. This is essentially 2-term
recursion with an initial guess for the electric field, 
which can be corrected afterwards
to yield the overall potential gain.
The recursion relation is
$\phi_{j+1} = 2 \phi_j - \phi_{j-1} - \rho_j \Delta x^2 /\epsilon_0$.
With smoothing, however, it becomes
$\phi_{j+1} = (2+\Delta x^2/ r_{\rm sm}^2) \phi_j 
   - \phi_{j-1} - \rho_j \Delta x^2 /\epsilon_0$,
which is unstable \citep{NumericalRecipes}; therefore, in 1D the 
smoothing cannot 
necessarily be done in the same way as the (unmodified) Poisson solve.  
On the other hand, if fast Fourier transforms are used for the
Poisson solve, then smoothing can be accomplished with almost 
no extra cost.
In 2D and 3D, where solution by direct integration is not possible,
the Poisson solve requires more complicated
matrix solves, and increased diagonal dominance may improve their
performance.

We will demonstrate in the next two sections that 
this modified Poisson smoothing can be used effectively to eliminate grid instability at all wavelengths.

\section{Grid instability growth rates for a Cauchy-squared distribution}
\label{sec:dispersionResults}

\begin{figure}
\centering 
\includegraphics*[width=3.in]{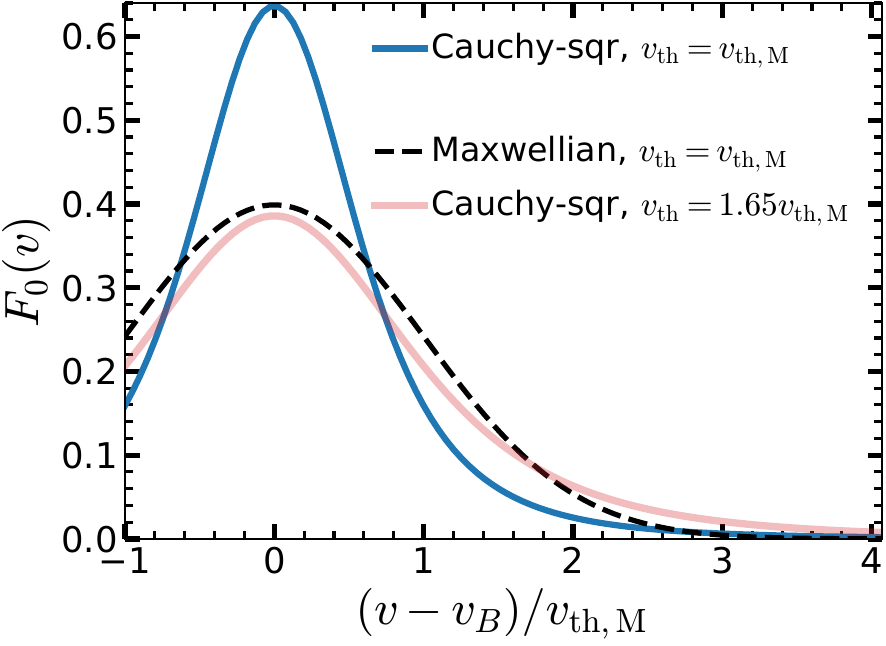}%
\caption{ \label{fig:CauchySqrDist}
A Cauchy-squared distribution $F_0(v)$ (blue line) with drift (beam)
speed $v_B$ and the same thermal 
speed $v_{\rm th}\equiv \sqrt{\langle (v-v_B)^2 \rangle}=v_{\rm th,M}$
as a Maxwellian (dashed), and a Cauchy-squared distribution
with $v_{\rm th}=1.65v_{\rm th,M}$ (pink), which 
has the same maximum slope as the Maxwellian.
}
\end{figure}

In this section we calculate the growth/decay rates (and
oscillation frequencies) of MC-PIC modes, in the limit of
zero time step and infinitely many macroparticles (but finite $\Delta x$),
by solving
the numerical dispersion $D(k,\omega)=0$ from Eq.~(\ref{eq:dispersion}), 
for the specific case of a Cauchy-squared distribution:
\begin{eqnarray}
  \label{eq:CauchySqrDist}
  F_0(v) &=& \frac{2 v_{\rm th}^3}{
    \pi [v_{\rm th}^2 + (v - v_B)^2]^2} \\
  \int F_0(v) dv &=& 1 \\
  \int v F_0(v) dv &=& v_B \\
  \label{eq:CauchySqrDistVth}
  \int (v-v_B)^2 F_0(v) dv &=& v_{\rm th}^2
\end{eqnarray}
where the drift/beam velocity $v_B$ and the
thermal velocity $v_{\rm th}$ are given 
in terms of the first and second moments.
This distribution roughly resembles a Maxwellian (see Fig.~\ref{fig:CauchySqrDist}),
but simplifies numerical calculation of the dispersion roots;
for reference the exact ($\Delta x\rightarrow 0$)
dispersion for the Cauchy-squared distribution
is solved in~Appendix~\ref{sec:dispersionExact}.
(A Cauchy distribution, 
$F_0(v) \propto 1/[v_{\rm th}^2 + (v-v_B)^2]$, leads to an even simpler
dispersion, but its velocity moments are inconveniently infinite.)
For simplicity, we will carry on as if $F_0(v)$ were 
interchangeable with a
Maxwellian with the same drift and thermal speeds, hence we define
$\lambda_D \equiv v_{\rm th}/\omega_p$, a choice that
yields that same Bohm-Gross correction to the plasma frequency as for a Maxwellian with the same $v_{\rm th}$ (Appendix~\ref{sec:dispersionExact}).
However, 
as Fig.~\ref{fig:CauchySqrDist} shows, 
a Cauchy-squared distribution 
more strongly resembles a Maxwellian
for $|v-v_B|\lesssim v_{\rm th}$ when
its thermal speed $v_{\rm th}$ is 50\%--65\% higher than the Maxwellian's.

Therefore, when attempting to compare behavior with a Maxwellian, 
$v_{\rm th}$ might need to be adjusted, probably by less than a factor
of~2.  This uncertainty should be of little practical concern, since most
PIC simulations probably cannot be trusted to contain a single 
perfect Maxwellian anyway.

With the Cauchy-squared distribution, we can approximate the numerical
dispersion as a finite-order polynomial with analytically-determined
error bounds over any given disk in the complex $\omega$ plane.  
The error can be made arbitrarily small by increasing the 
polynomial degree or decreasing the disk radius.
By using many overlapping disks, each with its own polynomial approximation 
of $D(k,\omega)$ for which roots must be found,
we can cover an area sufficient to find growth rates with desired accuracy 
($10^{-6}\omega_p$) for all physical modes and all growing modes.
Details of this process are given in Appendix~\ref{sec:dispersionSolving}.

Through nondimensionalization, the normalized complex mode frequencies 
$\omega/\omega_p$ [for which $D(k,\omega)=0$] are functions of the three 
dimensionless
parameters $k\Delta x$, $v_B/\omega_p \Delta x$, and
$v_{\rm th}/\omega_p \Delta x = \lambda_D/\Delta x$.
We explored a 3D parameter space to characterize instability,
numerically calculating the dispersion
over an exhaustive set of parameters:
\begin{itemize}[leftmargin=*,itemindent=-0.75in]
  \item
$k\Delta x/\pi \in \{0.00001$, 
$\pm 0.00002$, $\pm 0.00005$, $\pm 0.0001$, $\pm 0.0002$, 
$\pm 0.0005$, $\pm 0.001$, 
$\pm 0.002$, $\pm 0.005$, $\pm 0.01$, $\pm 0.02$, $\pm 0.05$, 
$\pm 0.1$, $\pm 0.2$, $\pm 0.3$, $\pm 0.4$, $\pm 0.5$, $\pm 0.6$,
$\pm 0.7$, $\pm 0.8$, $\pm 0.9$, $\pm 0.95$, 
$\pm 0.99\}$,
  \item
$v_B/\omega_p \Delta x \in \{0.0001$, $0.0002$, $0.0005$, $0.001$,
  $0.002$, $0.005$, $0.01$, $0.02$, $0.05$,
  $0.07$, $0.08$, $0.1$, $0.12$, $0.14$, $0.16$, $0.18$, 
  $0.2$, $0.22$, $0.24$, $0.26$, $0.28$, $0.30$, $0.32$, $0.34$, 
  $0.36$, $0.38$, $0.4$, $0.5$, $1\}$,
  \item
$\lambda_D/\Delta x \in \{0.0001$, 
$0.0005$, $0.001$, $0.002$, $0.005$, $0.01$, $0.02$, $0.05$, $0.07$, 
$0.08$, $0.1$, $0.12$, $0.14$, $0.16$, $0.18$, $0.2$, $0.22$, $0.24$, 
$0.26$, $0.28$, $0.3$, $0.32$, 
$0.34$, $0.36$, $0.38$, $0.4$, $0.5$, $0.7$, $1\}$.
\end{itemize}
We explored the effect of smoothing by considering 
\begin{eqnarray} \label{eq:simSmoothing}
  K(k)^2 &=& K_{\rm Poisson}(k)^2 [1 + K_{\rm Poisson}(k)^2 r_{\rm sm}^2]
     \\
     && \textrm{where } \frac{r_{\rm sm}}{\Delta x}
     = \frac{\alpha}{\pi} \frac{\Delta x}{\lambda_D}
     \nonumber
\end{eqnarray}
for ``stability factor'' $\alpha=0$ (standard MC-PIC without smoothing), 
as well as smoothed MC-PIC with
$\alpha=0.2$, $1$, 
and~$5$---cf. Eq.~(\ref{eq:stabilityCriterion}).
We used the standard 3-point finite-difference Poisson solve, not a
Fourier solve---see the definitions of $K^2$ and $\kappa$ after 
Eq.~(\ref{eq:dispersion}).

\begin{figure*}
\centering 
\newlength{\wfig}
\setlength{\wfig}{3.1in}
\includegraphics*[width=\wfig]{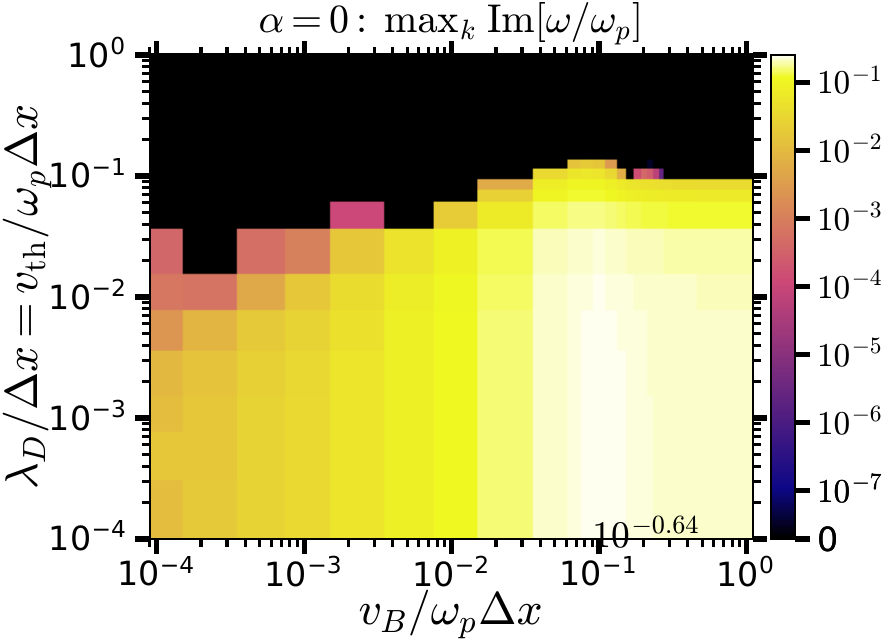}%
\raisesubfiglabel{0.1in}{\wfig}{(a)}%
\includegraphics*[width=\wfig]{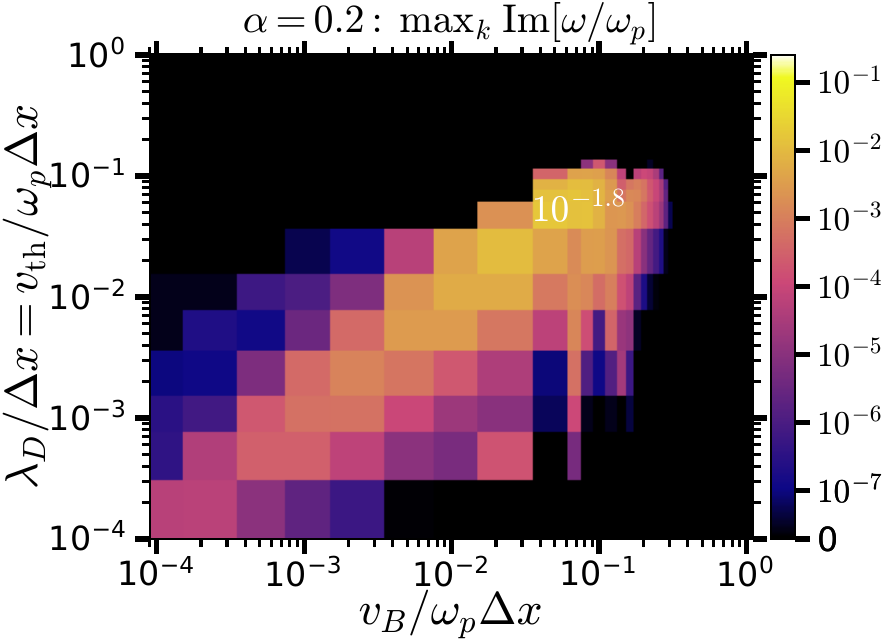}%
\raisesubfiglabel{0.1in}{\wfig}{(b)}
\\
\includegraphics*[width=\wfig]{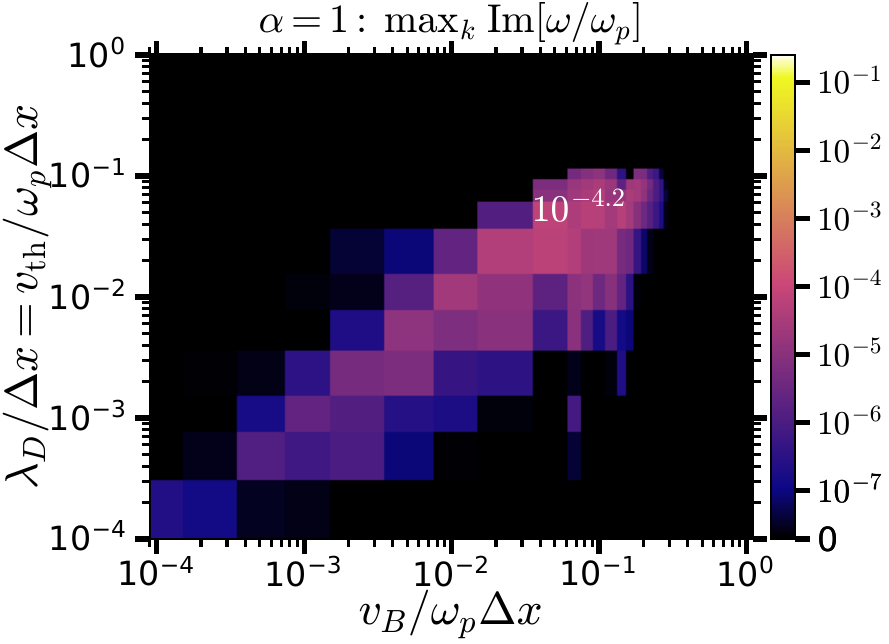}%
\raisesubfiglabel{0.1in}{\wfig}{(c)}%
\hspace{0.1in}%
\includegraphics*[width=\wfig]{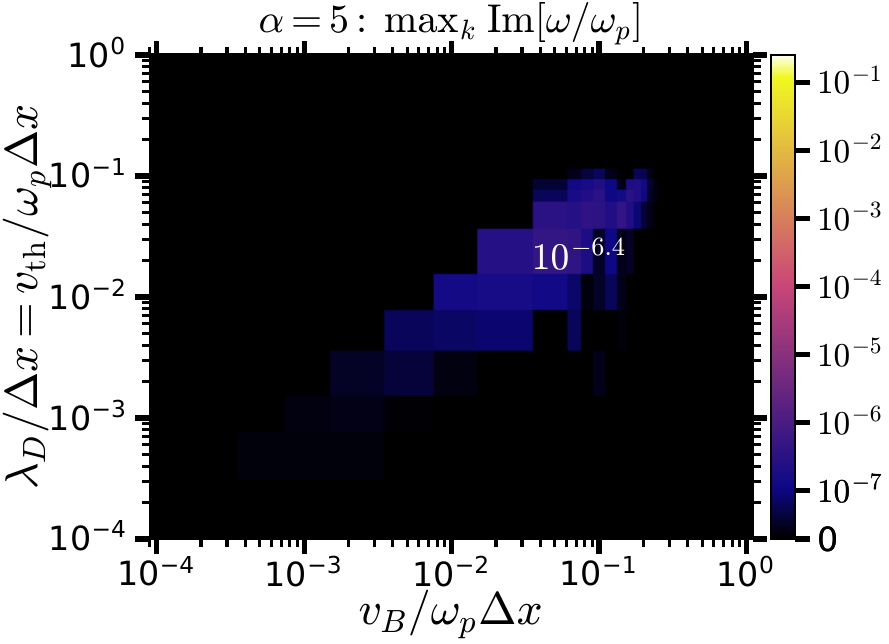}%
\raisesubfiglabel{0.1in}{\wfig}{(d)}
\caption{\label{fig:growthRate}
Increasing the smoothing length 
$r_{\rm sm} = (\alpha/\pi)\Delta x^2 / \lambda_D$,
for smoothing with Eq.~(\ref{eq:simSmoothing}), stabilizes grid heating.
The panels show, for $\alpha=0$ (standard PIC), 0.2, 1, and 5,
the maximum growth rate (over all $k$), normalized to the plasma
frequency, for MC-PIC simulation of uniform electron plasma initialized
with a Cauchy-squared distribution, Eq.~(\ref{eq:CauchySqrDist}),
that has normalized 
drift velocity $v_B/\omega_p \Delta x$ and thermal velocity
$v_{\rm th}/\omega_p \Delta x = \lambda_D/\Delta x$.
The maximum value over each plot is labeled on each plot.
Measurements below $10^{-6}\omega_p$ are possibly consistent with zero.
}
\end{figure*}

By solving the numerical dispersion, 
we found the maximum growth rates for MC-PIC plasma modes.
Figure~\ref{fig:growthRate} shows the maximum growth rate (over all~$k$),
normalized to the plasma frequency, as a function of 
$v_B/\omega_p\Delta x$ and $\lambda_D/\Delta x$.
Growth rates below $10^{-6}\omega_p$ may be below the error of the
calculation, and hence may be consistent with zero.
With no smoothing (Fig.~\ref{fig:growthRate}a, $\alpha=0$, i.e., standard MC-PIC), simulations with $\lambda_D/\Delta x \gtrsim 0.14$ 
are stable and those with $\lambda_D/\Delta x \lesssim 0.02$ are
highly unstable, regardless of~$v_B$. 
For $v_B/\omega_p \Delta x \gtrsim 10^{-2}$, the transition 
from stability to
strong instability is very rapid; 
for $\lambda_D/\Delta x$ just 20\% below instability threshold,
the growth rate is within an order of magnitude of the 
maximum observed growth rate of about $0.22\omega_p$.

In the limit of large $v_B$, the instability threshold (without smoothing) 
approaches a constant that can be calculated analytically, 
$\lambda_D/\Delta x \approx 0.092$ (which agrees with 
Fig.~\ref{fig:growthRate});
in this limit we can neglect all but one 
term (one value of $g$) in the sum in Eq.~(\ref{eq:dispersion}), 
yielding a cubic
equation that can be solved in the same way as the exact dispersion
(see Appendix~\ref{sec:dispersionExact}).
For a drifting Maxwellian plasma, 
\citet{Birdsall_Maron-1980} find, experimentally,
that the instability threshold is $\lambda_D/\Delta x \approx 0.046$
(in the large $v_B$ limit).  We speculate that it is not 
$v_{\rm th}^2\equiv \langle (v-v_B)^2 \rangle$ 
that determines grid instability in this case.  
Instead, the maximum 
slope of $F_0(v)$ may be the critical factor, since
it determines the Landau damping rate (for weak damping).
In this case, a Cauchy-squared distribution with 
$v_{\rm th}=1.65 v_{\rm th,M}$ is ``equivalent'' to a Maxwellian
with $v_{\rm th}=v_{\rm th,M}$ (Fig.~\ref{fig:CauchySqrDist}).
Defining $\lambda_{D,M}\equiv v_{\rm th,M}/\omega_p$, we would
expect the Cauchy-squared simulation to be stable (in the large $v_B$
limit) for $\lambda_D/\Delta x = 1.65 \lambda_{D,M}/\Delta x > 0.092$,
or $\lambda_{D,M}/\Delta x > 0.056$, very close to  
the value of 0.046.

For $v_B \rightarrow 0$ (without smoothing),
we find (looking at Fig.~\ref{fig:growthRate}, 
$\alpha=0$) that the instability threshold is roughly
$\lambda_D/\Delta x \approx 0.04$ for 
$v_B/\omega_p \Delta x \sim 10^{-4}$, but we do
not know how to extrapolate this to $v_B=0$.  This threshold 
is significantly
lower than the estimate $\lambda_D/\Delta x \approx 0.15$ from running 
MC-PIC simulations with Maxwellian distributions \citep{Adams_etal-2025inPrep}.
We do not have a good explanation for this difference.
In any case, the exact threshold value is only of theoretical 
interest,
except possibly for simulations with extremely high~$M_{\rm ppc}$;
most simulations will be ruined by particle noise if they run
anywhere near the grid instability threshold.

Increasing the smoothing to $\alpha=0.2$ 
(Fig.~\ref{fig:growthRate}b), i.e., smoothing over
$r_{\rm sm}/\Delta x=0.06 \Delta x/\lambda_D$ cells,
significantly reduces instability 
growth rates, stabilizing 
simulations with $v_B/\omega_p \Delta x> 0.3$, but the maximum
growth rate of $0.016 \omega_p$ is likely still too high for 
most simulations.
A further increase to $\alpha=1$ (Fig.~\ref{fig:growthRate}c), or 
$r_{\rm sm}/\Delta x=0.3\Delta x/\lambda_D$,  
reduces the growth rates by another 2 orders
(maximum: $6\times 10^{-5} \omega_p$).
Using $\alpha=2$ (not shown, 
$r_{\rm sm}/\Delta x=0.6\Delta x/\lambda_D$)
reduces the maximum growth rate to
$6\times 10^{-6}\omega_p$, and $\alpha=5$ 
(Fig.~\ref{fig:growthRate}d, $r_{\rm sm}/\Delta x=1.6\Delta x/\lambda_D$) reduces it
below $10^{-6}\omega_p$ (we measure
$4\times 10^{-7}\omega_p$; however, since our numerical calculation may 
have errors as high as $10^{-6}\omega_p$, 
this may be consistent with zero).

Thus smoothing suppresses grid instability.
Our prediction in~\S\ref{sec:dispersion} 
was that if $\lambda_D/\Delta x\gtrsim \alpha_c \approx 0.14$
ensures
stability without smoothing, as Fig.~\ref{fig:growthRate}(a) suggests, 
then smoothing with $\alpha=\alpha_c=0.14$ 
would marginally stabilize the Nyquist mode, $k\Delta x=\pi$.
This appears to be the case, but stabilizing $k=\pi/\Delta x$ is not
sufficient: modes with smaller $k$ must also be stabilized, and the
smoothing affects them less than the Nyquist mode.

As we increase the smoothing length~$r_{\rm sm}$, the
fastest growing modes shift to smaller~$k$.
Almost all the fastest growing modes for $\alpha=0$ (no smoothing, 
standard MC-PIC)
have $|k_{\rm fastest}|\Delta x > 0.7\pi$
(for comparison, using a Fourier Poisson-solve results in the fastest modes 
having $|k_{\rm fastest}|\Delta x \approx \pi$; when we use the 3-point finite-difference Poisson solve, the subsequent interpolation of
edge $E$ to nodal $E$ yields $\kappa(\pi/\Delta x)=0$,
eliminating the Nyquist mode).
For $\alpha=0.2$, 
$|k_{\rm fastest}| \Delta x < 0.6\pi$; 
for $\alpha=1$, $|k_{\rm fastest}|\Delta x<0.1\pi$, and
for $\alpha=5$, $|k_{\rm fastest}|\Delta x<0.02\pi$.
(In this last case, our coverage of $k$ becomes pretty sparse
around $k_{\rm fastest}$, so we could potentially miss some more unstable
modes; however, MC-PIC simulation in the following section backs up these
results.)
We speculate that for $\alpha \lesssim 0.14$, 
$|k_{\rm fastest}|\Delta x \sim \pi$ and for $\alpha \gtrsim 0.14$, 
$|k_{\rm fastest} \Delta x| \sim 0.14\pi/\alpha$.
In achieving stability, 
our smoothing [i.e., a low-pass filter of $\rho$ with Fourier profile
$(1+k^2r_{\rm sm}^2)^{-1}$] overstabilizes the shortest-wavelength modes.
There may be more optimal filters that achieve stability with less
degradation of the spatial resolution;
we leave such questions to future work.

We have thus shown how smoothing suppresses grid instabilities for
a Cauchy-squared distribution.
We have also characterized grid instability growth rates, as a
function of $v_B/\omega_p \Delta x$ and $\lambda_D/\Delta x$,
for standard MC-PIC (with a Cauchy-squared distribution)
without smoothing.
Although grid instability is absent for $\lambda_D/\Delta x \gtrsim 0.14$
(for any $v_B$),
under-resolving the
Debye length (without smoothing)
can still lead to problems with particle noise,
which is absent in the preceding treatment, 
since it assumes an infinite number of particles.
We will investigate the effect of smoothing on grid instability and 
PIC noise using full 1D MC-PIC simulation in the following section.

\section{Results from 1D MC-PIC simulations}
\label{sec:PICresults}

Although solving the numerical dispersion 
is a powerful tool for understanding
PIC simulation, it does not include the effects of finite time step or
a finite number of particles.
Therefore we test smoothing directly with a simple 1D MC-PIC simulation,
and investigate how PIC noise depends on $\lambda_D/\Delta x$
(as well as other parameters, including $M_{\rm ppc}$ and $r_{\rm sm}$).

\subsection{MC-PIC simulation set-up}
\label{subsec:setup}

We implemented the standard MC-PIC algorithm: 
leapfrogged momentum-conserving electrostatic PIC 
\citep[with linear weighting, sometimes called area weighting or
cloud-in-cell;][]{BirdsallPIC}
with one spatial and one velocity dimension, using a 3-point
finite-difference Poisson solve, coded in Python.
We used a regular grid with $N$ cells and cell size $\Delta x$, 
hence length $L=N \Delta x$, and periodic boundary conditions.

We initialized electrons with $M_{\rm ppc}$ 
macroparticles per cell, 
evenly distributed
to yield zero initial electric field
(except in~\S\ref{subsec:noisyStart}, where 
we explore random initial positions, leading to non-zero random
initial electric field).
The field solve assumed a uniform compensating
positive charge.
Each electron was given a velocity drawn from a random 1D Maxwellian 
distribution (not a Cauchy-squared distribution as used above)
with drift velocity $v_B$ and thermal velocity $v_{\rm th}$.
We note that the actual average velocity differs 
from $v_B$ 
by order $v_{\rm th}/\sqrt{N M_{\rm ppc}}$
due to statistical variation.
(However, in~\S\ref{subsec:noisyStart}, we explore ``quiet'' starts with
particles initially ordered in phase space.)

We applied smoothing (at each time step) with strength $\alpha$ as described 
in~\S\ref{sec:poissonSmoothing}; i.e., the smoothing radius is
$r_{\rm sm}/\Delta x=(\alpha/\pi)\Delta x/\lambda_D$ cells.

We varied the time step $\Delta t$ from simulation to simulation, always
keeping it below the maximum stable value.
Given a maximum plasma mode frequency
$\omega_{\rm max}$, the leapfrog time advance requires 
$\omega_{\rm max} \Delta t<2$ for stability
(although $\omega_{\rm max} \Delta t \ll 1$ is required for accuracy at frequency $\omega_{\rm max}$).
Often the highest frequency is simply $\omega_p$; however, with
a drifting plasma, $\omega = kv_B \pm  \omega_p$, possibly 
requiring a smaller time step.
In addition, it is important for stability and accuracy that most particles
travel less than $\Delta x$ in one 
time step \citep{Brackbill_Forslund-1982}.  Since 
$v_{\rm th} = \lambda_D \omega_p$, 
the stability criterion 
$(\pi v_B/\Delta x + \omega_p) \Delta t < 2$ means that
$v_{\rm th}\Delta t/\Delta x < 2/(\pi v_B/v_{\rm th} + \Delta x/\lambda_D)$
and so whenever $\lambda_D/\Delta x\ll 1$, the cell-crossing criterion
is always met; for $\lambda_D/\Delta x \gtrsim 1$, we have to reduce
the time step.
We ran simulations for a time $T$, which we will usually express
in terms of the plasma period $2\pi/\omega_p$; we ran some simulations
up to $T=2\pi 10^4/\omega_p$.

We carried out all simulations with 
$\lambda_D=1\:$m, $\omega_p=1\:$s, hence $v_{\rm th}=1\:$m/s,
and varied $\Delta x$, $L$, $v_B$, $M_{\rm ppc}$, $r_{\rm sm}$,
and $\Delta t$.
Thus we explored stability and noise over the
6-dimensional parameter space of dimensionless parameters
$v_B/\omega_p \Delta x$, 
$\lambda_D/\Delta x=v_{\rm th}/\omega_p \Delta x$, 
$M_{\rm ppc}$, $\alpha=r_{\rm sm}/\Delta x$,
$\omega_p \Delta t$,
and $N=L/\Delta x$.
Our results scale trivially to any choice of $\lambda_D$ and $\omega_p$.

When we discuss results below, we will specify the values of these
parameters.  We will hardly mention the effect of $N$, which is negligible
as long as $N\gg 1$ and $r_{\rm sm}/\Delta x \ll N$.
Similarly, decreasing the time step had negligible effect on the results
reported here.

\subsection{Signatures of grid instability and noise-driven heating}
\label{subsec:signatures}

\begin{figure*}
\centering 
\makebox[0in]{\raisebox{1.395in}[0in][0in]{\hspace{4.5in}%
  \includegraphics*[width=2.11in,trim=0 4.01in 0 0]{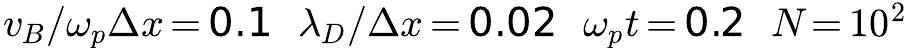}%
}}
\includegraphics*[width=2.11in,trim=0 0 0 0.33in]{fig3a.pdf}%
\raisesubfiglabel{0.08in}{2.11in}{(a)}%
\hspace{0.04in}%
\includegraphics*[width=2.11in,trim=0 0 0 0.33in]{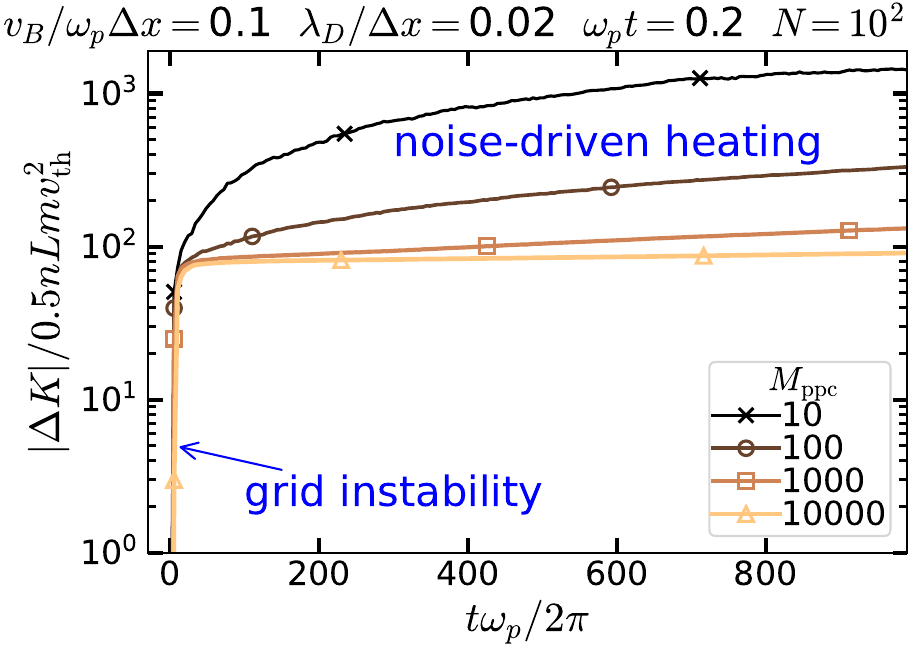}%
\raisesubfiglabel{0.08in}{2.11in}{(b)}%
\hspace{0.04in}%
\includegraphics*[width=2.05in]{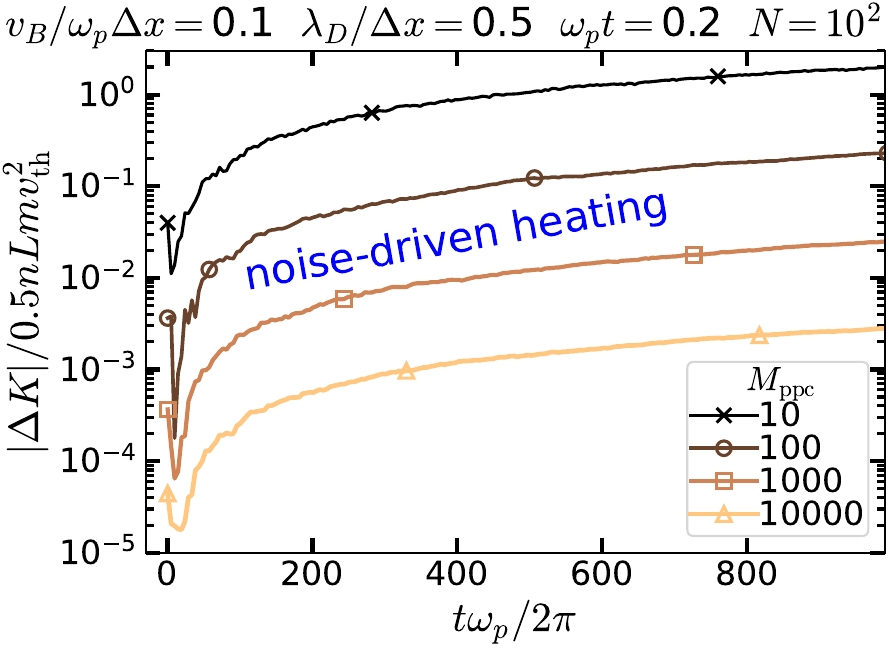}
\raisesubfiglabel{0.08in}{2.05in}{(c)}%
\caption{\label{fig:vB0p1}
Grid instability and noise-driven heating have different
$M_{\rm ppc}$-dependence.
Each panel shows the magnitude of plasma kinetic energy change over time,
normalized to the initial thermal energy, 
for MC-PIC simulations with no smoothing ($\alpha=0$) and $M_{\rm ppc}$
ranging from 10 to $10^4$.
Panels (a) and (b) show the same simulation with 
$\lambda_D/\Delta x=0.02$ (and $v_B/\omega_p \Delta x=0.1$).
(a)~In the first few plasma periods, 
grid instability causes exponential energy growth; 
the effect of $M_{\rm ppc}$ on the growth rate and saturation energy
diminishes for large~$M_{\rm ppc}$.
(b)~Over longer times, particle noise causes additional unphysical
heating.
(c)~A simulation with
$\lambda_D/\Delta x=0.5$ (and $v_B/\omega_p \Delta x=0.1$)
exhibits no grid instability, but does suffer from noise-driven heating,
which scales as $M_{\rm ppc}^{-1}$.
}
\end{figure*}

We start by showing MC-PIC simulation without smoothing.
Figure~\ref{fig:vB0p1} shows the time evolution of (the absolute
value of) the net change in kinetic 
energy~$\Delta K(t)$ for
two cases, one with $\lambda_D/\Delta x=0.02$ that suffers 
grid instability and one 
with $\lambda_D/\Delta x=0.5$ that does not.
Both cases have $v_B/\omega_p \Delta x=0.1$, $N=100$ cells, and
time step $\omega_p \Delta t=0.2$; in each case, we
show four simulations with different~$M_{\rm ppc}$.
In the first few plasma periods of the unstable case (Fig.~\ref{fig:vB0p1}a) 
$\Delta K(t)$ grows exponentially; 
the electric field energy grows at
about the same rate (not shown).  As $M_{\rm ppc}$ increases, the 
exponential growth in energy ($\sim e^{2 \gamma t}$) appears to converge to
rate $\gamma \approx 0.2 \omega_p$, roughly as we calculated 
in Fig.~\ref{fig:growthRate} (for a slightly different 
velocity distribution~$F_0$).
This exponential growth saturates at a level nearly independent of 
$M_{\rm ppc}$,
about 60 times the initial plasma energy (while the electric field energy
grows to about 10 times the initial plasma energy).
Increasing the energy by 60 times increases 
$\lambda_D/\Delta x$ to about $0.02\sqrt{60}\approx 0.15$, which is 
roughly where
we expect stability (albeit for a Cauchy-squared distribution,
Fig.~\ref{fig:growthRate}; this agreement is probably 
coincidentally closer than
the uncertainties of this estimate).

However, over longer times (Fig.~\ref{fig:vB0p1}b, showing $|\Delta K(t)|$ for the same simulation out to 1000 plasma periods) the energy continues to grow unphysically at a
rate that depends strongly on $M_{\rm ppc}$.
With $M_{\rm ppc}=10^4$, the additional heating over 1000 plasma periods
is noticeable but small compared with the grid instability heating.
For $M_{\rm ppc}=10$, however, 
the noise-driven heating increases the energy by another
order of magnitude.

In Fig.~\ref{fig:vB0p1}(c), we show $|\Delta K|$ 
over time in a marginally-resolved case
with $\lambda_D/\Delta x=0.5$, 
again for different $M_{\rm ppc}$.
Immediately after initialization, 
the electrons lose some energy to the electric field,
which is initially zero (cf.~\S\ref{subsec:velocityDiffusion}); thus $\Delta K$ is negative at very
early times, and
then grows, rising through zero and becoming large and positive.
None of these simulations show an exponential growth phase; 
grid instability is absent.
Nevertheless, the energy grows unphysically due to stochastic 
noise-driven
heating. The
growth is strongly $M_{\rm ppc}$-dependent:
$\Delta K(t) \propto M_{\rm ppc}^{-1}$.
For $M_{\rm ppc}=10$ heating is significant
after just a few tens of plasma periods, but for $M_{\rm ppc}=1000$, 
$\Delta K$
grows to only a couple percent of the initial kinetic energy over 
1000 periods.
The energy grows roughly linearly in time, consistent
with particles exhibiting a random walk in velocity space.

Thus we distinguish grid instability heating
from noise-driven (stochastic) heating.
Grid instability causes exponential growth
independent of $M_{\rm ppc}$ for sufficiently large~$M_{\rm ppc}$.
It saturates when the Debye length increases to 
$\lambda_D/\Delta x \sim 0.1$.
However, with $\lambda_D/\Delta x \lesssim 1$, 
particle noise is enhanced (cf.~\S\ref{sec:noiseTheory}). 
Field fluctuations 
due to statistical particle noise (cf.~\S\ref{subsec:Enoise})
cause diffusion in velocity space,
leading to unphysical heating that scales as~$M_{\rm ppc}^{-1}$
(and grows linearly in time; cf.~\S\ref{subsec:velocityDiffusion}).

\subsection{Smoothing suppresses grid instability and noise-driven heating}
\label{subsec:suppression}

\begin{figure*}
\centering 
\includegraphics*[width=3.1in]{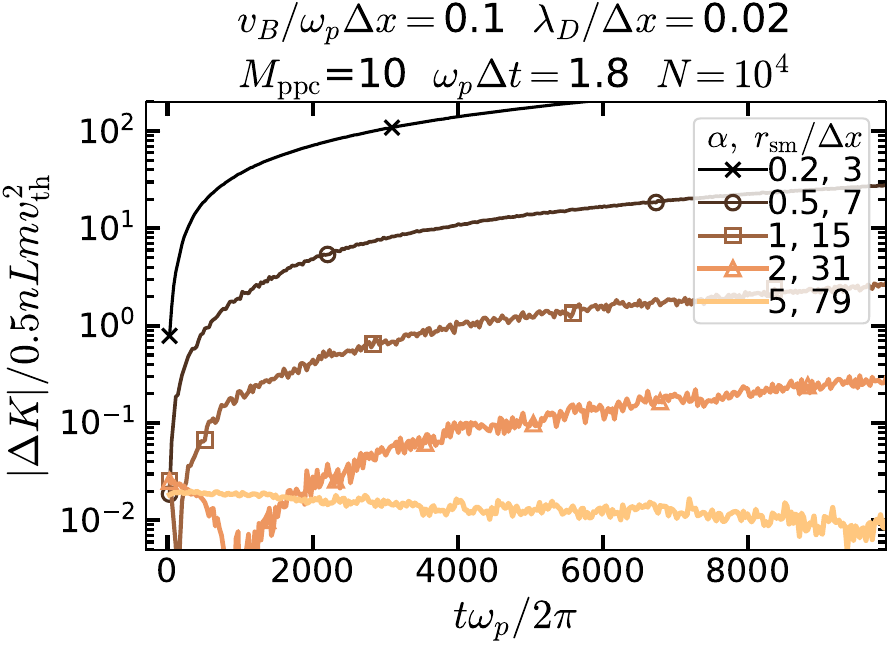}%
\raisesubfiglabel{0.1in}{3.1in}{(a)}%
\includegraphics*[width=3.1in]{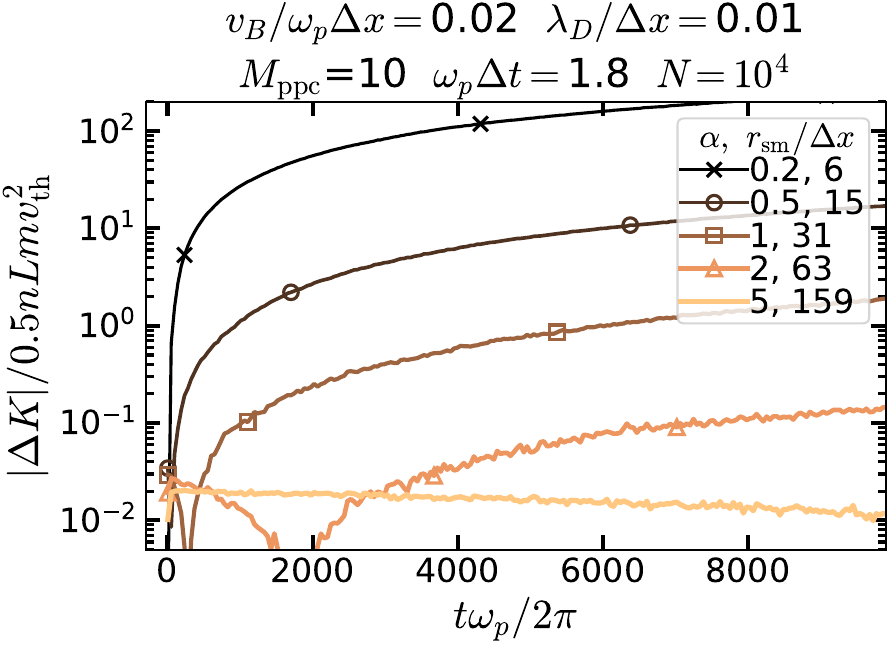}%
\raisesubfiglabel{0.1in}{3.1in}{(b)}
\\
\includegraphics*[width=3.1in]{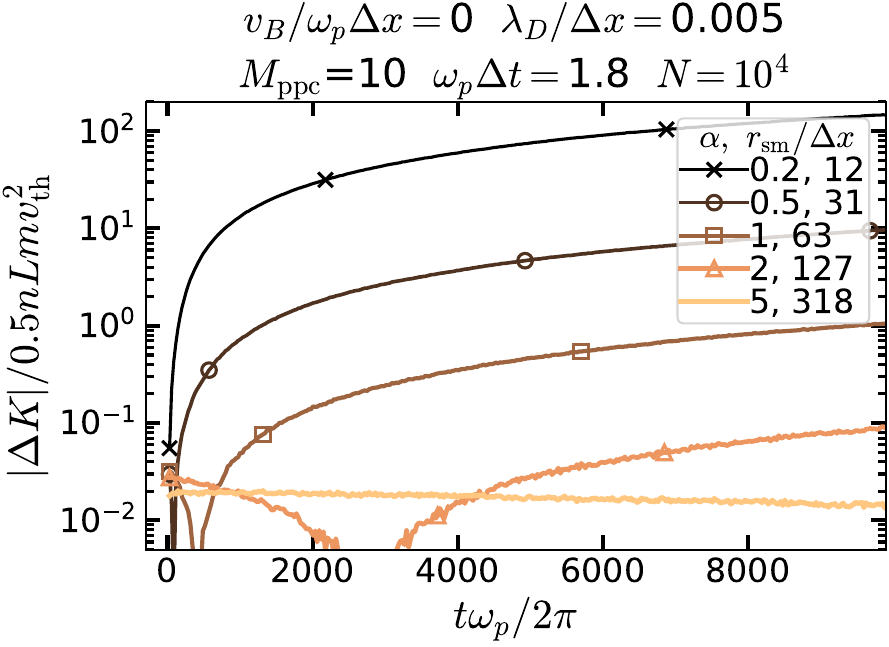}%
\raisesubfiglabel{0.1in}{3.1in}{(c)}%
\includegraphics*[width=3.1in]{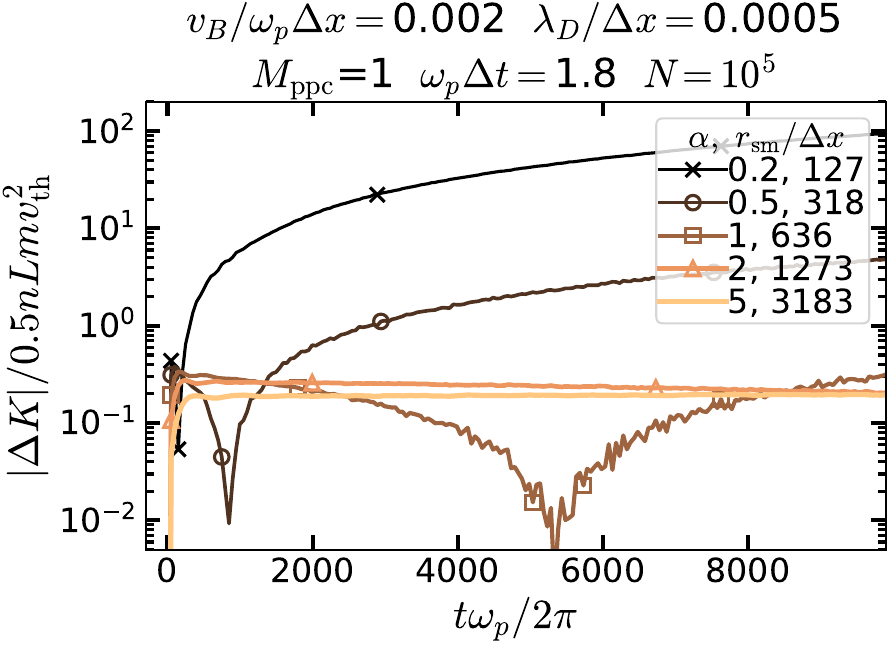}%
\raisesubfiglabel{0.1in}{3.1in}{(d)}
\caption{\label{fig:smoothedStable}
Smoothing suppresses grid instability (which is so strong 
without smoothing that it would just be a spike on the left side of these
graphs) and noise-driven heating.
These plots show the magnitude of plasma kinetic energy change 
$|\Delta K|$
over $10^4$ plasma periods,
for four 1D MC-PIC simulations with different $v_B/\omega_p \Delta x$ 
and $\lambda_D/\Delta x$ (labeled at the top of each panel, along 
with $M_{\rm ppc}$, the time step, and number of cells), 
each with different amounts of
smoothing: $\alpha=0.2$, 0.5, 1, 2, and 5 (the corresponding 
$r_{\rm sm}/\Delta x$ is labeled for each~$\alpha$).
While $\alpha=1$ might be tolerable (especially for shorter times),
$\alpha=5$ provides robust simulation for long times.
In the severely under-resolved case with $\lambda_D/\Delta x=0.0005$
(panel d), only $M_{\rm ppc}=1$ is used, and the resulting noise
causes $|\Delta K|$ to be 
immediately $\sim$20\% of the initial energy for $\alpha \gtrsim 1$;
this can be reduced to about 2\% by using $M_{\rm ppc}=10$ (as seen in
the other panels).
}
\end{figure*}

We now demonstrate that smoothing can eliminate grid instability in
actual PIC simulation, confirming the results 
of~\S\ref{sec:dispersionResults} (but for Maxwellian distributions); 
we will also see that smoothing suppresses noise-driven heating.
Figure~\ref{fig:smoothedStable} shows the kinetic energy versus time
for four very under-resolved cases
[with $(v_B/\omega_p \Delta x, \lambda_D/\Delta x)$ equal to
$(0,0.005)$, $(0.02,0.01)$, $(0.1,0.02)$, and $(0.002,0.0005)$]. 
In each case, we show the simulations with different smoothing lengths,
$\alpha = 0.2$, 0.5, 1, 2, and 5 (also labeling the smoothing radius in
cells, $r_{\rm sm}/\Delta x$).
They all run for $10^4$ plasma periods.
Three have $N=10^4$ and $M_{\rm ppc}=10$, and the
one with worst Debye resolution has $N=10^5$ (so that $r_{\rm sm}\ll L$) 
and $M_{\rm ppc}=1$.
The time step, $\omega_p \Delta t=1.8$, is large---too large
for accurate measurement of plasma frequencies,
but it shows that, despite $v_B$, one can use a time step 
nearly up to $\omega_p \Delta t = 2$.
A smaller time step yields similar stability
and should render Langmuir waves accurately if their wavelength
is long compared to~$r_{\rm sm}$.

All these simulations are highly unstable for $\alpha=0$ (not shown), 
with unphysical
energy gains of multiple orders of magnitude over tens of plasma periods.
A little smoothing ($\alpha =0.2$) reduces unphysical heating 
substantially, taking thousands of plasma periods to increase the energy
by a disastrous factor of~100.
With $\alpha=1$, the unphysical energy 
gain over $10^4$ plasma periods is of the order of the initial energy
(or several times that, when $M_{\rm ppc}=1$).
With $\alpha=2$, however, simulation is reasonable, at least for
short times (hundreds or thousands of plasma periods), 
and $\alpha=5$ is sufficient to avoid unphysical heating for
$10^4$ plasma periods.
With $\alpha=5$, the 
particle distribution after $10^4$ plasma periods
looks essentially identical
to the initial distribution (with $10^6$ particles in the simulation,
the distributions can be compared out to velocities 
about $3.9v_{\rm th}$ from the mean); 
for the $(0.02,0.01)$ case, we measured 
the same $v_B/\omega_p \Delta x$ to high precision (because the algorithm
is momentum conserving), while $v_{\rm th}/\omega_p \Delta x$ changed
by $-0.6\%$.

Even for $\alpha=0.2$, the energy growth, hence the need for
large $r_{\rm sm}$,
appears to be driven primarily by noise, not grid instability (cf. Fig.~\ref{fig:vB0p1}).  
Interestingly, with sufficient smoothing, the noise level no longer
depends on $r_{\rm sm}$; this is especially clear for $M_{\rm ppc}=1$,
where $|\Delta K|$ is roughly the same for $\alpha=2$ and $\alpha=5$.
Moreover, for $\alpha =5$, 
$|\Delta K| \approx 0.2 M_{\rm ppc}^{-1}$ for all 4 cases.  
(This is discussed more in~\S\ref{subsec:velocityDiffusion}.)

Smoothing also allows large time steps  
$\omega_p \Delta t \approx 2$, even for large drift velocities
(probably because high-$k$ modes are suppressed by smoothing).
With explicit time-stepping, we cannot exceed
$\omega_p \Delta t = 2$ 
because plasma oscillations still have frequency
$\omega_p$ at the longest wavelengths (we have observed problems
for $\omega_p \Delta t$ just slightly less than 2, but as we show above,
$\omega_p \Delta t = 1.8$ works well).

Because $\lambda_D/\Delta x$ is so small in these cases,
smoothing with $\alpha=5$ requires large smoothing radii
$r_{\rm sm}$ of
80, 160, 320, and 3200 cells in the four cases shown in Fig.~\ref{fig:smoothedStable}. 
Moreover, $r_{\rm sm}$ is the characteristic exponential decay length of the smoothing kernel, which thus extends somewhat beyond
$r_{\rm sm}$.

The stabilizing of extremely under-resolved simulations with
$\lambda_D/\Delta x=5 \times 10^{-4}$
and $M_{\rm ppc}=1$ dramatically confirms the effectiveness of this
approach.  
However, smoothing over even $r_{\rm sm}/\Delta x=80$
cells seems repugnant at first because it significantly degrades the
spatial resolution;
nevertheless, as we will discuss in~\S\ref{sec:discussion}, this
would be done in a situation where the desired resolution is
$\Delta x_{\rm ideal} \sim 80^2 \lambda_D$, and such smoothing
enables the use of $\Delta x\sim 80\lambda_D$, a cell size 80
times larger than unsmoothed MC-PIC could use, 
likely speeding up simulation
by more than a factor of 80.

\subsection{Electric field noise}
\label{subsec:Enoise}

It is particularly important to understand
the effects of noise when under-resolving the Debye length.
In this section we characterize the electric field
noise in MC-PIC simulations with and 
without smoothing. 
For brevity, we focus mostly on cases with zero drift velocity, $v_B=0$.

When each macroparticle represents many electrons, randomness in
macroparticle positions can lead to unphysical electric field fluctuations.
After initializing a simulation with evenly-spaced (perfectly correlated) 
macroparticles yielding $E=0$, the root mean square electric field 
variation, $E_{\rm rms}$, increases over time but remains much less than
$E_{f\Delta x}$ in 
Eq.~(\ref{eq:randomNoise}) with $f\sim M_{\rm ppc}^{-1/2}$,
which applies to completely uncorrelated particle positions.
Thus particles become less correlated but far from uncorrelated.

\begin{figure*}
\centering
\includegraphics*[width=3.1in]{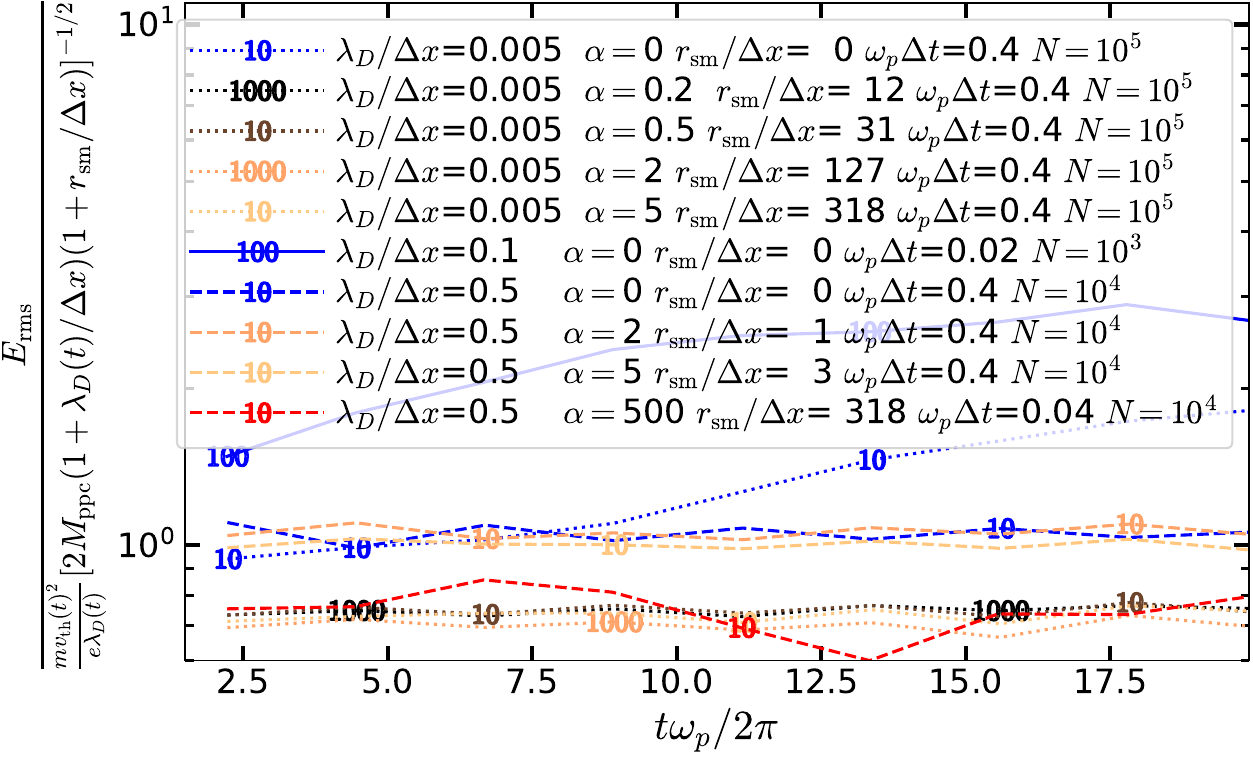}%
\raisesubfiglabel{0.05in}{3.1in}{(a)}%
\hspace{0.08in}%
\includegraphics*[width=3.1in]{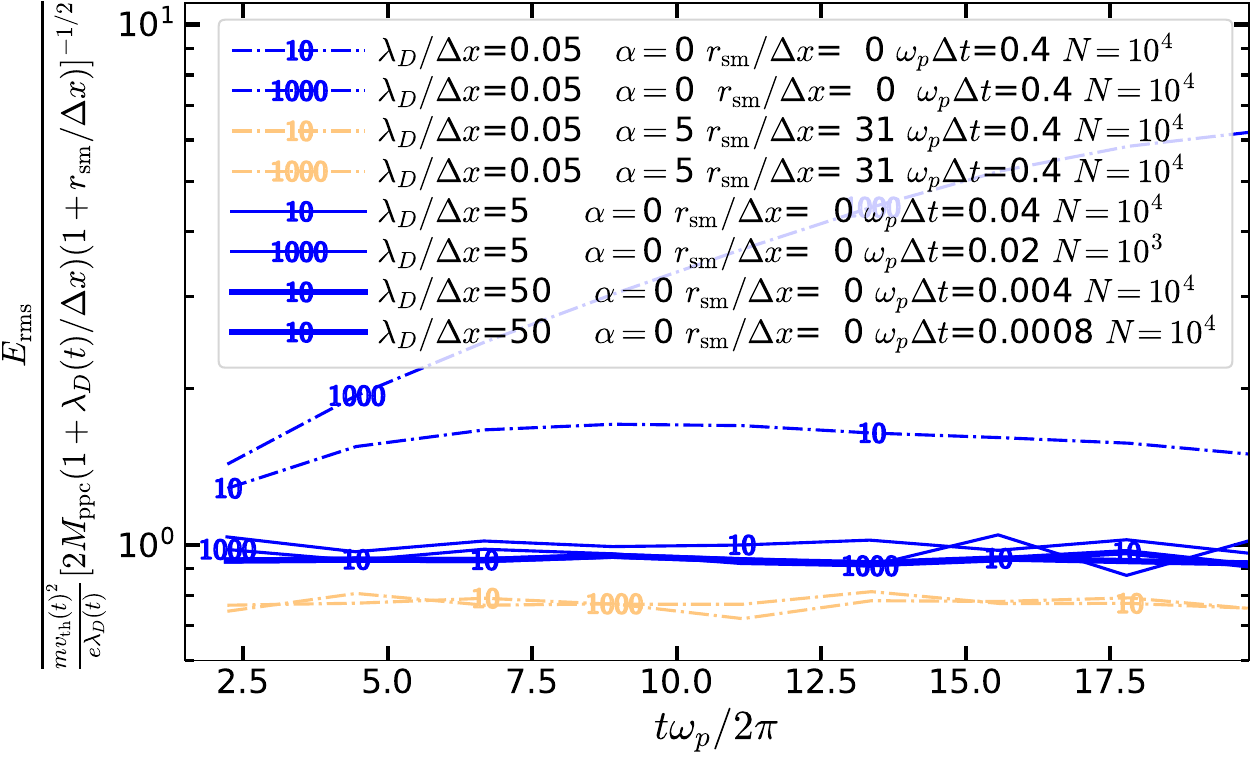}%
\raisesubfiglabel{0.05in}{3.1in}{(b)}
\caption{\label{fig:Erms}
The point of these plots is merely that all the values are of order one, confirming that Eq.~(\ref{eq:Erms}) is a decent estimate of the ``noise field'' $E_{\rm rms}$ for a wide range of $\lambda_D/\Delta x$, $M_{\rm ppc}$, $\alpha$, 
$\omega_p \Delta t$, and $N$.
The two panels  
both show $E_{\rm rms}$ normalized to Eq.~(\ref{eq:Erms}) over 20 plasma
periods for different sets of simulations with $v_B=0$.
The varied parameters are labeled in the legend, except for 
$M_{\rm ppc}\in\{10,100,1000\}$,
which is written on the curves.
We make no attempt to explain the variations beyond
Eq.~(\ref{eq:Erms}).
Here, the normalization depends on time through $v_{\rm th}(t)$ 
and $\lambda_D(t)$,
which may be significant for
simulations with $\lambda_D/\Delta x \ll 1$ and no smoothing.
}
\end{figure*}

We find empirically (at least for $v_B=0$)
that the root-mean-square electric field is
\begin{eqnarray} \label{eq:Erms}
  E_{\rm rms} &\sim& \frac{m v_{\rm th}^2}{e\lambda_D}
    \frac{1}{\sqrt{2 M_{\rm ppc} 
      (1+\lambda_D/\Delta x)(1+r_{\rm sm}/\Delta x)}}
\end{eqnarray}
where $v_{\rm th}$ and $\lambda_D$ are immediate 
values (not initial values).
This is demonstrated in Fig.~\ref{fig:Erms}, which shows $E_{\rm rms}$
calculated over the whole simulation length at regular times over the
first 20 plasma periods, for $v_B=0$.
The measured values of $E_{\rm rms}$ over this time are almost all within
a factor of 2 of Eq.~(\ref{eq:Erms}),
for parameters ranging in 
$0.005 \leq \lambda_D/\Delta x \leq 50$, 
$10 \leq M_{\rm ppc} \leq 10^3$,
$0.0008 \leq \omega_p \Delta t \leq 0.4$,
$0 \leq \alpha \leq 5$,
$10^4 \leq N \leq 10^5$.
In 2 cases ($\lambda_D/\Delta x=0.05$, $\alpha=0$ and $0.2$), 
$E_{\rm rms}$ evolved over time to be about 2.5 times and
about 6 times higher than Eq.~\ref{eq:Erms} after 20 plasma periods.
Unlike $E_{f\Delta x}$ in Eq.~(\ref{eq:Erms}), the self-consistent
$E_{\rm rms}$ depends only weakly on $\lambda_D/\Delta x$ for
$\lambda_D < \Delta x$.

In terms of energy density
(making use of $v_{\rm th}=\lambda_D \omega_p$),
\begin{eqnarray}\label{eq:ErmsEnergy}
  \frac{\epsilon_0}{2} E_{\rm rms}^2 &\sim& 
    \frac{n m v_{\rm th}^2}{4 M_{\rm ppc} 
      (1+\lambda_D/\Delta x)(1+r_{\rm sm}/\Delta x)}
\end{eqnarray}
suggesting that simulations with $\lambda_D\lesssim \Delta x$
and $r_{\rm sm} = 0$
are near {\it numerical} 
equipartition with the field energy per cell equal to
the thermal energy of one macroparticle,
$(\epsilon_0/2)E^2 = n m v_{\rm th}^2/(2 M_{\rm ppc})$
\citep[we speculate that the factor of 2 discrepancy here might be related to 
the particle shape spanning two cells, which provides a small amount of smoothing even when $r_{\rm sm}=0$;][]{Hockney-1971}.
This agrees well with analytical predictions of
\citet{Touati_etal-2022} for 1D PIC with no smoothing, 
$(\epsilon_0/2) E_{\rm rms}^2 \sim 
   (1/4)
   (n m v_{\rm th}^2/M_{\rm ppc}) (2 \Delta x/ \pi \lambda_D)
   \tan^{-1}(\pi \lambda_D/2 \Delta x)$.
I.e., $\tan^{-1}(x)/x \approx (1 + 2x/\pi)^{-1}$ within 25\% error, which
is less than the scatter of our measurements.
Smoothing or resolving $\lambda_D$ decreases $E_{\rm rms}$
below numerical equipartition with macroparticles.

\subsection{Quiet and noisy starts}
\label{subsec:noisyStart}

\begin{figure*}
\centering 
\includegraphics*[height=0.5in]{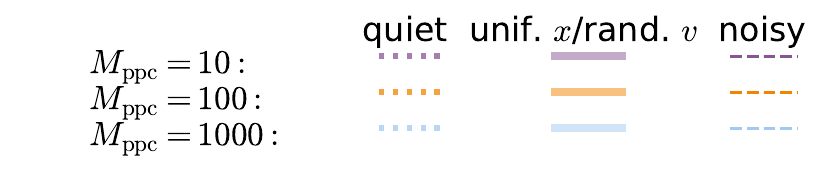}\\
\includegraphics*[width=2.1in]{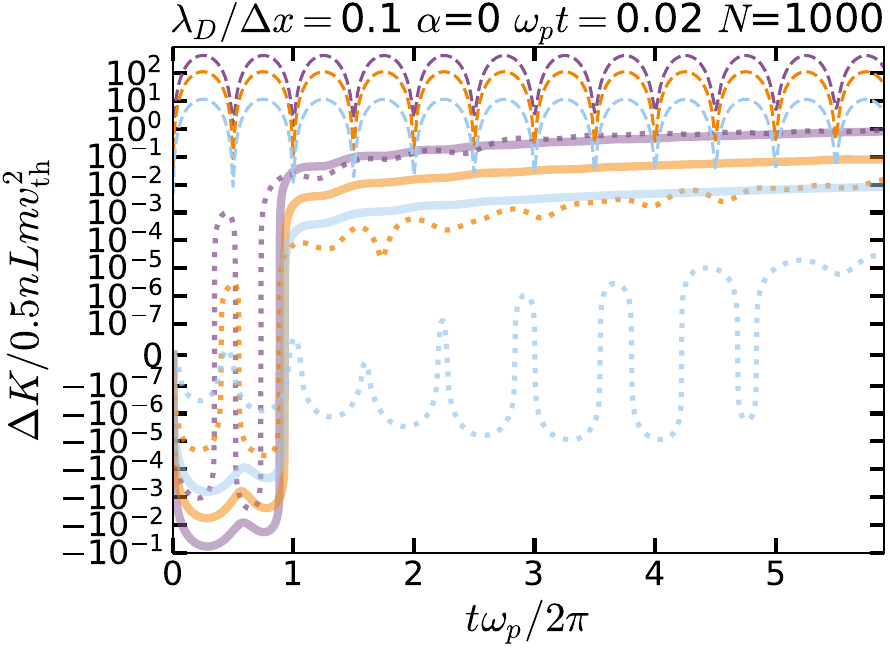}%
\raisesubfiglabel{0.08in}{2.1in}{(a)}%
\hspace{0.08in}%
\includegraphics*[width=2.1in]{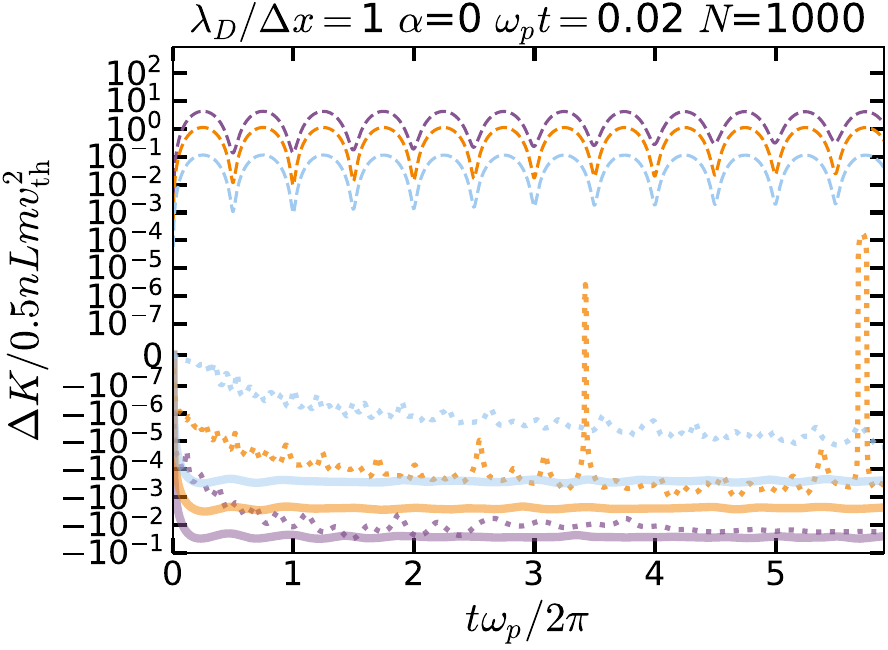}%
\raisesubfiglabel{0.08in}{2.1in}{(b)}%
\hspace{0.08in}%
\includegraphics*[width=2.1in]{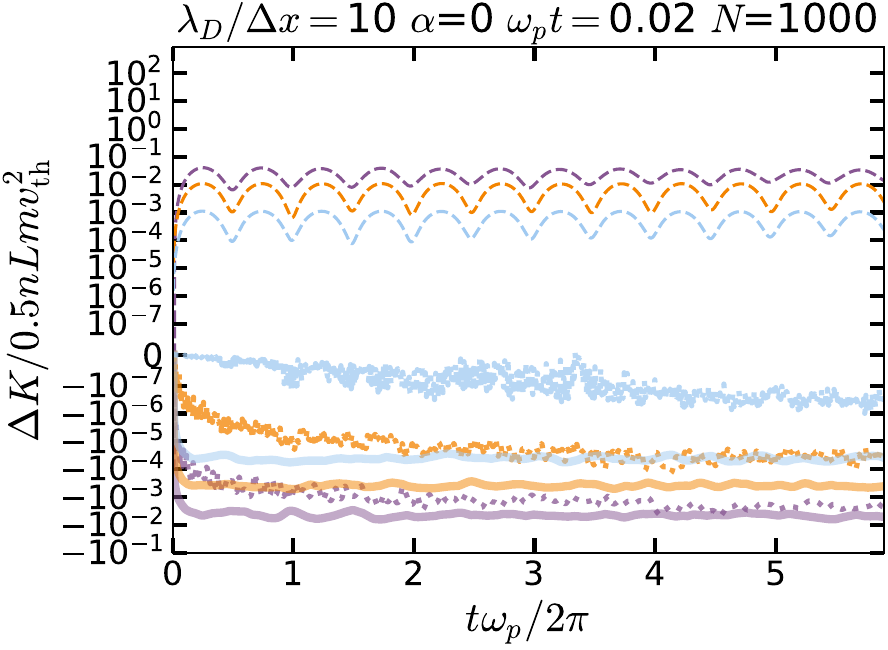}%
\raisesubfiglabel{0.08in}{2.1in}{(c)}
\caption{\label{fig:quietVsNoisyStart}
The (unphysical)
change $\Delta K$ in total particle energy over six plasma periods,
for different particle initialization strategies,
shown for  $\lambda_D/\Delta x=0.1$, $1$, and $10$ (left to right),
and for different $M_{\rm ppc} \in \{10, 100,1000\}$ (purple, orange,
light blue).
Dotted lines show a quiet start with ordered positions and velocities,
yielding the smallest $|\Delta K|$ for given $\lambda_D/\Delta x$ 
and~$M_{\rm ppc}$.
Thick, solid lines show ordered positions with random velocities
(used throughout this paper)---this does nearly as well as a quiet start 
for small~$M_{\rm ppc}$.
Thin, dashed lines show a noisy start with random positions and velocities;
the random positions immediately create a large electric field
(cf.~\S\ref{sec:noiseTheory}) that sloshes to the particles at the
plasma frequency.
When starting with ordered positions, hence $E=0$, particles initially
lose energy to the electric field; a quiet start (at least with 
$\lambda_D/\Delta x \gtrsim 1$) prolongs this phase.
For $\lambda_D /\Delta x =0.1$, noise-driven heating starts after
just one plasma period for ordered positions with random velocities;
a quiet start can delay this too, and reduce noise-driven heating,
although eventually (even in this uniform plasma) 
its initial correlations will be lost.
All simulations have $v_B=0$, $\omega_p t=0.02$, and no smoothing.
}
\end{figure*}

Since a uniform plasma (with an infinite number of particles)
would have $E=0$, $E_{\rm rms}$ is the ``noise'' field that depends
on how interactions between particles correlate their 
positions.  Anything that
places particles in positions via mechanisms other than the PIC 
algorithm may yield different
noise levels.  For example, initializing a simulation 
by placing particles at {\it random} positions will yield 
$E_{\rm rms} \sim E_{f\Delta x}$ with $f\approx M_{\rm ppc}^{-1/2}$
from Eq.~(\ref{eq:randomNoise}).

Figure~\ref{fig:quietVsNoisyStart}
shows the change in plasma energy $\Delta K$ over the 
first six plasma periods, for three different initialization strategies,
for
$\lambda_D/\Delta x \in \{0.1,1,10\}$ and $M_{\rm ppc}\in\{10,100,1000\}$.
A ``noisy'' start (dashed lines), with {\it random} initial
positions and velocities, immediately introduces a relatively
large unphysical electric field $\sim E_{f\Delta x}$
\citep[also see][]{Acciarri_etal-2024a,Acciarri_etal-2024b}; the field 
energy is transferred to particles via conspicuous plasma oscillations.
For $\lambda_D/\Delta x=0.1$ (Fig.~\ref{fig:quietVsNoisyStart}a), even 
$M_{\rm ppc}=10^3$ 
introduces an unphysical energy ten times the initial plasma energy.
For $\lambda_D/\Delta x=1$ (Fig.~\ref{fig:quietVsNoisyStart}b), $M_{\rm ppc}=10^2$ still
yields 100\% unphysical energy gain (this falls to 10\% for
$M_{\rm ppc}=10^3$).
In contrast, 
for $\lambda_D/\Delta x=10$ (Fig.~\ref{fig:quietVsNoisyStart}c),
a noisy start with $M_{\rm ppc} > 10$ yields an initial field energy
less than a few percent of the plasma energy.

The thick lines show the result of initially {\it uniformly-spaced}
positions
with {\it random} velocities
(like most simulations in this paper).  Here, $\Delta K$ is 
negative at early times, 
because the electric field is initially zero and immediately 
gains some energy from particles (this is discussed more 
in~\S\ref{subsec:velocityDiffusion}).

For comparison, we also show a ``quiet'' start (dotted lines), 
initialized with equal-weight particles  
{\it ``evenly'' spaced in both position and velocity}
\citep{Byers_Grewal-1970,BirdsallPIC}.
The generated velocities were
spaced within intervals along the velocity axis such that 
every interval had the same area under the velocity distribution $f(v)$.
(In the same equal-area sense, each velocity was located at 
the ``middle'' of its interval.)
The velocities, at first ordered monotonically,
were then scrambled via a bit-reversal permutation.
Even quieter starts are possible,
in particular using ordered velocities in one cell and copying
the particles exactly into the other cells, but such high order can
introduce new instabilities \citep{Gitomer_Adam-1976}.
In any case, the low-noise correlations will be lost over time, especially
with small $M_{\rm ppc}$.
For small $M_{\rm ppc}\sim 10$, the difference between 
random and ordered velocities can be pretty insignificant.
Even for $M_{\rm ppc}=100$, 
after only 20 plasma periods (not shown)
the unphysical energy with a quiet start 
is just a factor of 3 lower than with uniform positions and random 
velocities.

Incidentally, grid instability occurs for
$\lambda_D/\Delta x = 0.1$ in Figure~\ref{fig:quietVsNoisyStart}(a),
but it is essentially undetectable because it is swamped 
by noise-driven heating, even for $M_{\rm ppc}= 1000$ with a quiet
start.  Using much higher $M_{\rm ppc}$ would 
reveal the exponential energy growth due to grid instability.

\subsection{Thermal energy evolution and velocity-diffusion heating}
\label{subsec:velocityDiffusion}

In our MC-PIC
simulations of uniform plasma that do not exhibit grid instability
(with initially evenly-spaced positions and random velocities),
we observe the following evolution of the plasma thermal energy,
which should ideally remain constant:
(we note that with momentum-conserving PIC, 
the bulk drift energy does remain constant)
\begin{itemize}
  \item Initially, the plasma has thermal energy density 
      $U_i=n m v_{{\rm th},i}^2/2$.
  \item Stage 1. 
    Almost immediately (in less than a plasma period)
    the plasma loses energy density
    $\epsilon_0 E_{\rm rms}^2/2$ [cf. Eq.~(\ref{eq:ErmsEnergy})] to the
    electric field, roughly conserving energy overall.
    For $\lambda_D \lesssim \Delta x$ and $r_{\rm sm}=0$,
    the plasma and field thus reach numerical equipartition immediately.
  \item Stage 2 (applicable only for $r_{\rm sm}>\Delta x$).
    Over time 
    $t_{\rm decl} \sim (1+r_{\rm sm}/\Delta x) / (\sqrt{2} \omega_p)$,
    the thermal energy density declines further, linearly in time,
    until it reaches
    $U_{\rm min} \sim U_i (1 - 0.3/M_{\rm ppc})$.
    If the electric field gained this energy, the system
    would near numerical equipartition; that does not happen, however,
    and this change violates energy conservation.
    With strong smoothing, we have observed $U_{\rm min}$ closer to
    $U_i(1-0.2/M_{\rm ppc})$; unfortunately,
    smoothing has little effect on $U_{\rm min}$,
    although it does lengthen $t_{\rm decl}$.
  \item Stage 3 (noise-driven heating). 
    The thermal energy density grows  linearly in time
    due to stochastic velocity diffusion (for which we find 
    an empirical formula below).
    The electric field energy also grows correspondingly, 
    violating energy conservation.
\end{itemize}
(N.B. We have not investigated the regime where both 
$\lambda_D > \Delta x$ and $r_{\rm sm}>0$.)
We find empirically that
the rate of change of the plasma thermal energy 
$K_{\rm th}$ is given in these three stages roughly by
\begin{widetext}
\begin{eqnarray} \label{eq:dKthdt}
  \frac{1}{\omega_p K_{\rm th}(t)}
  \frac{dK_{\rm th}}{dt}(t)
  &\sim &
  \left\{ \begin{array}{l@{\quad\textrm{for }}l}
    -\frac{\delta(\omega_p t)}{2 M_{\rm ppc} 
      (1+\lambda_D/\Delta x)(1+r_{\rm sm}/\Delta x)}
      & \omega_p t \approx 0 \\
    -\frac{0.3 \sqrt{2}}{M_{\rm ppc}(1+r_{\rm sm}/\Delta x)}
    \phantom{\frac{\displaystyle 1^1}{\displaystyle 1^1}}
      & 0 \lesssim \omega_p t \lesssim 
                  \frac{1+r_{\rm sm}/\Delta x}{ \sqrt{2}}
                  \\
  \frac{ 5 \times 10^{-4}  
    \left( \frac{\Delta x}{\lambda_D(t)} \right)^3 }{
    M_{\rm ppc}
    \left[ 1 + \left(\frac{r_{\rm sm}}{\Delta x}\right)^2 \right]
    \left( 1 + \frac{r_{\rm sm}}{\lambda_D(t)} \right)
  }
           & \frac{1+r_{\rm sm}/\Delta x}{ \sqrt{2}} 
                 \lesssim  \omega_p t 
  \end{array} \right.
\end{eqnarray}
\end{widetext}
where we use the Dirac delta function $\delta(\omega_p t)$ to reflect 
the rapid equilibration of electric field energy with plasma on 
short time scales that we have not tried to measure.
In the third stage, noise-driven heating,
$dK_{\rm th}/dt$ is constant (as long as $\Delta K \ll K_{\rm th}$); thus $K_{\rm th}$ grows linearly in time, consistent with diffusion in velocity.
In all three stages, $\Delta K \propto M_{\rm ppc}^{-1}$;
thus particle weight can play an important role in unphysical 
energy exchanges \citep[see, e.g.,][]{Lawson_Gray-1991,Jubin_etal-2024}.

\begin{figure*}
\centering 
\includegraphics*[width=2.13in]{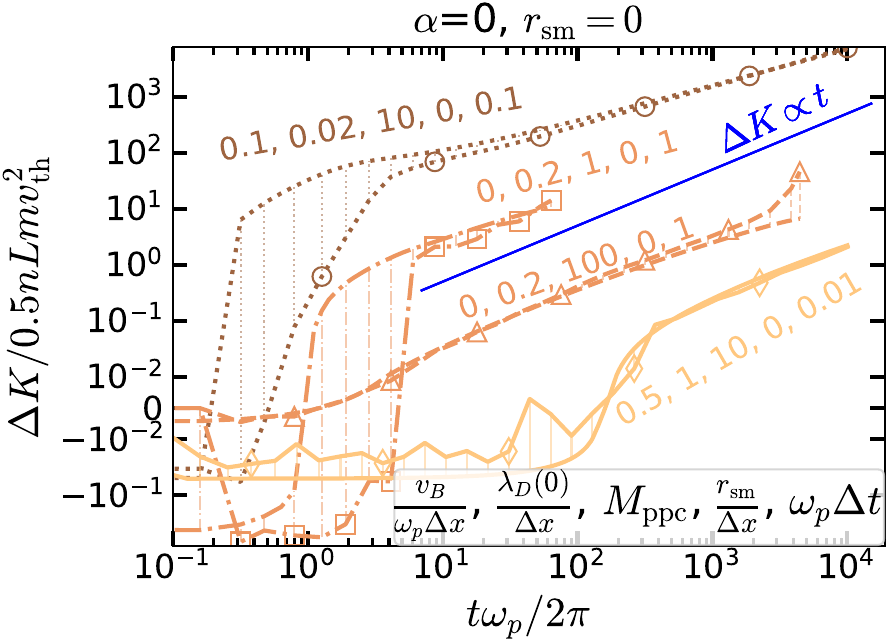}%
\raisesubfiglabel{0.05in}{2.13in}{(a)}%
\hspace{0.03in}%
\includegraphics*[width=2.13in]{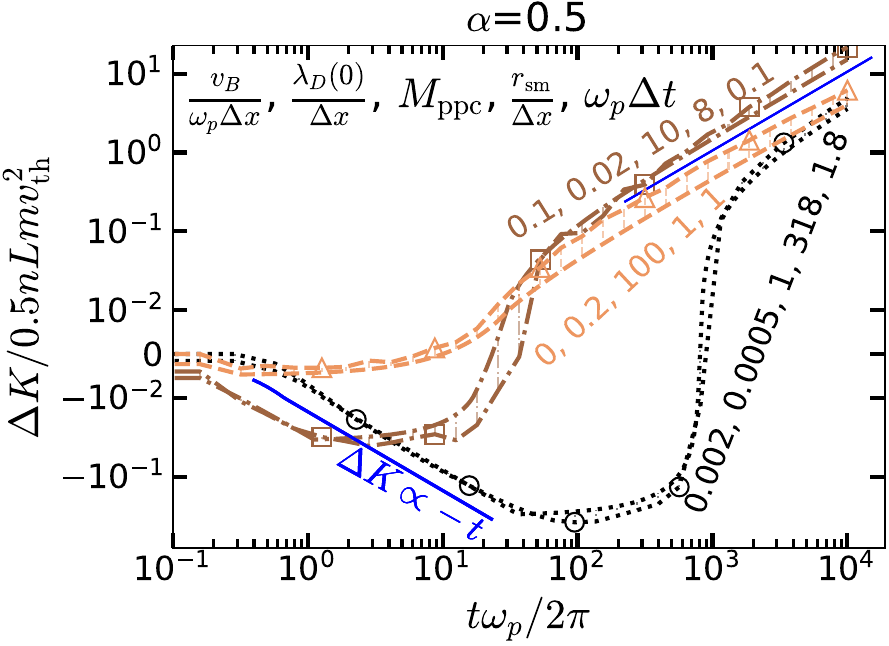}%
\raisesubfiglabel{0.05in}{2.13in}{(b)}%
\hspace{0.03in}%
\includegraphics*[width=2.13in]{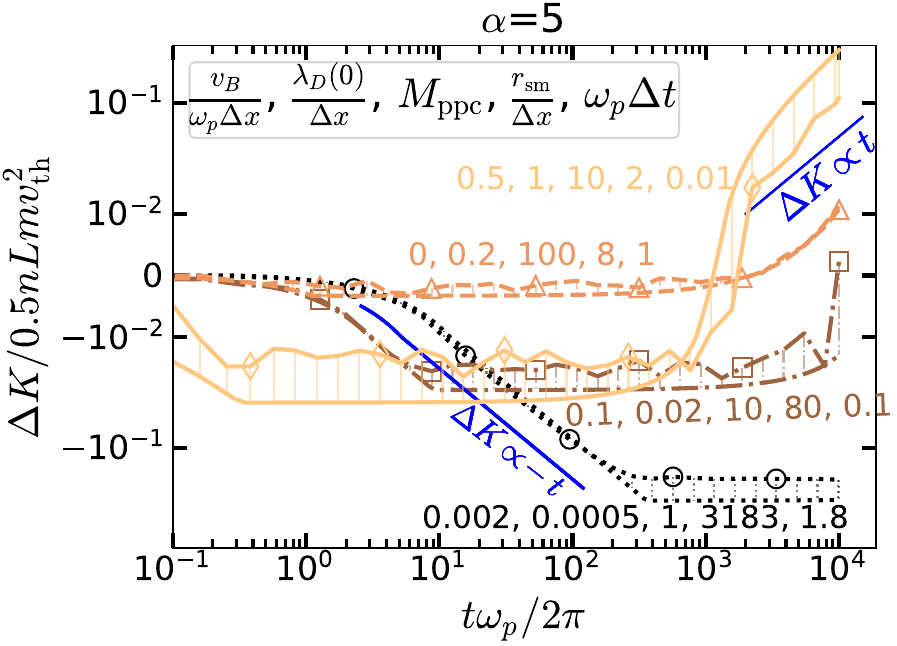}%
\raisesubfiglabel{0.05in}{2.13in}{(c)}%
\caption{\label{fig:dKth} 
Each pair of lines, connected by vertical segments,
shows the measurement and fit of (unphysical) net change in plasma energy,
$\Delta K$, versus time for MC-PIC simulations of uniform plasma.
For each pair, the line with symbols shows simulation data
and the line without symbols 
shows numerical integration of Eq.~(\ref{eq:dKthdt}).
Color indicates the value of $\lambda_D/\Delta x$ at $t=0$,
from dark [$\lambda_D(0)/\Delta x=0.0005$] 
to light [$\lambda_D(0)/\Delta x=1$].
For reference, solid blue lines show $\Delta K \propto \pm t$.
Aside from small time-shifts,
the data and fit are close for a wide range of simulation parameters
(see below).
In each case, we see the immediate loss in plasma energy 
($\Delta K < 0$ at $\omega_p t\ll 1$; stage 1); 
for $r_{\rm sm} > \Delta x$, energy loss 
continues as $\Delta K \propto -t$ for time $t_{\rm decl}$ (stage 2).  
Then (stage 3) the plasma energy starts increasing as $\Delta K \propto t$,
consistent with diffusion in velocity.
The three panels show simulations with different smoothing,
$\alpha=0$, $0.5$, and $5$, from (a) to (c), and in each panel,
simulations are labeled individually by $v_B/\omega_p \Delta x$, 
$\lambda_D(0)/\Delta x$, $M_{\rm ppc}$, 
$r_{\rm sm}/\Delta x$, and $\omega_p \Delta t$.
The simulations shown here include 
$v_B/\omega_p\Delta x \in \{0, 0.002, 0.1\}$, 
$\lambda_D(0)/\Delta x \in \{0.0005, 0.02, 0.2, 1\}$,
$M_{\rm ppc} \in \{1,10,100\}$,
$0 \leq r_{\rm sm}/\Delta x < 3200$, and
$0.01 \leq \omega_p \Delta t \leq 1.8$.
Simulations that end before $T=2\pi 10^4/\omega_p$ were halted because
more than 1/4 of the particles gained enough energy to cross $\Delta x$
in time $\Delta t$, which usually precedes a rapid energy rise
[briefly visible, e.g., for $t\omega_p/2\pi \gtrsim 3000$ 
in panel (a) ($\alpha=0$) orange/dashed
case with $v_B/\omega_p \Delta x=0$,
$\lambda_D(0)/\Delta x=0.2$, $M_{\rm ppc}=100$].
}
\end{figure*}

Figure~\ref{fig:dKth} shows the accuracy of this empirical formula for
11 simulations covering a range of 
$v_B$, initial $v_{\rm th}$ [i.e., $\lambda_D(0)$], 
$M_{\rm ppc}$, $r_{\rm sm}$, and $\Delta t$.
Each pair of lines, connected by thin vertical segments, shows the 
evolution of $\Delta K(t)$---the line with symbols is simulation data,
and the line without symbols is a numerical time-integration of
Eq.~(\ref{eq:dKthdt}).
The fit is good within a factor of a few, except for some small 
shifts in time.

In the first stage, $\omega_p t \lesssim 0.5$, some plasma energy is immediately lost to
the electric field as the initial state with perfectly-correlated
particle positions and $E=0$ gives way to a less-correlated 
(but far from uncorrelated) state with $E\sim E_{\rm rms}$
\citep[cf.][]{Acciarri_etal-2024a,Acciarri_etal-2024b}.
Accordingly, $\Delta K < 0$ at the earliest times shown in
Fig.~\ref{fig:dKth}; smoothing reduces this effect.
However, with smoothing ($r_{\rm sm} \gtrsim \Delta x$)
there is a second stage where
the plasma continues to lose energy linearly in time 
($\Delta K \propto -t$, paralleling the solid blue lines) over a longer
period $t_{\rm decl}$,
until it has lost a fraction
$\sim 0.3/M_{\rm ppc}$ of its original thermal energy.
After that (the third stage), 
$\Delta K$ starts growing linearly in time 
($\Delta K \propto t$,
parallel the ascending solid blue lines), consistent
with typical particle velocities diffusing as the square root of time.
When $\Delta K$ becomes comparable to $K_{\rm th}$, the diffusive
heating slows as $\lambda_D(t)$ becomes better resolved.

With velocity diffusion, and without smoothing ($r_{\rm sm}=0$), 
the energy of one physical electron within a macroparticle grows (during the third stage)
on average as
\begin{eqnarray} \label{eq:diffEnergyGrowth}
  \Delta K_1(t) &\sim & 2.5 \times 10^{-4} (\Delta x/\lambda_D)^3
    M_{\rm ppc}^{-1} m v_{\rm th}^2 \omega_p t
.\end{eqnarray}
It is interesting to compare the energy growth with a simplistic
random walk in velocity, where particle velocities 
receive a random kick 
$\Delta v = \pm eE_{\rm diff} \tau/m$ (kicks with $+$ and $-$
have equal probability) during each interval of time~$\tau$.
In this model (which depends on choice of $E_{\rm diff}$ and~$\tau$), 
\begin{eqnarray} \label{eq:randWalk}
  \frac{m}{2}\langle v(t)^2 \rangle &\sim & 
    \frac{m}{2} \frac{e^2 E_{\rm diff}^2 \tau^2}{m^2} \frac{t}{\tau} 
\end{eqnarray}
(where angle brackets are ensemble averages), and so comparison with
Eq.~(\ref{eq:diffEnergyGrowth}) yields
\begin{eqnarray}
  E_{\rm diff} \sqrt{\omega_p \tau} &\sim &
  \sqrt{5 \times 10^{-4}} \left( \frac{\Delta x}{\lambda_D}\right)^{3/2}
  \frac{m v_{\rm th}^2}{e\lambda_D} M_{\rm ppc}^{-1/2} 
\end{eqnarray}
If we speculate that $\omega_p \tau \sim 1$ (which, we will see in
the next subsection, works for test particles), then
$E_{\rm diff} \sim 0.03 (\Delta x/\lambda_D)^{3/2} 
\sqrt{1+\lambda_D/\Delta x} E_{\rm rms}$.
For $\lambda_D/ \Delta x \gtrsim 0.1$, $E_{\rm diff} < E_{\rm rms}$,
so velocity diffusion (especially at $\lambda_D/\Delta x \gg 0.1$)
is slower than would be expected from random $E_{\rm rms}$-sized 
kicks in every time interval~$\sim \omega_p^{-1}$.
It is reasonable to suppose that the effective $\tau$ depends on
$\lambda_D/\Delta x$, but we find this does not satisfactorily explain
the entire $\lambda_D/\Delta x$ dependence. 
Rather, we suspect that the correlations between particle positions
prevent particles from experiencing uncorrelated (over~$\tau$) 
fields of magnitude $E_{\rm rms}$ in the way that a test particle would.

Thus we have shown that unphysical energy growth in MC-PIC is consistent
with stochastic 
velocity diffusion, insofar as the energy grows linearly in time.
However, we have not produced a model that predicts $E_{\rm diff}$ 
and $\tau$ independently, although we have empirically determined
(assuming a simple random walk in velocity) the value and scaling of
$E_{\rm diff} \sqrt{\tau}$.

\subsection{Test particles do not always mimic MC-PIC particles}
\label{subsec:testParticles}

\begin{figure*}
\centering 
\includegraphics*[width=2.1in]{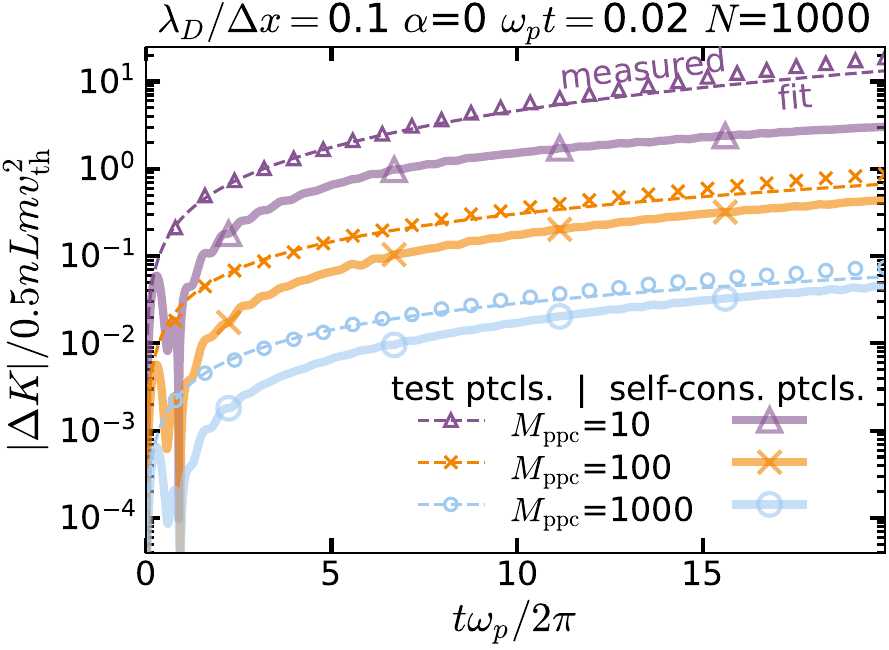}%
\raisesubfiglabel{0.05in}{2.1in}{(a)}%
\hspace{0.04in}%
\includegraphics*[width=2.1in]{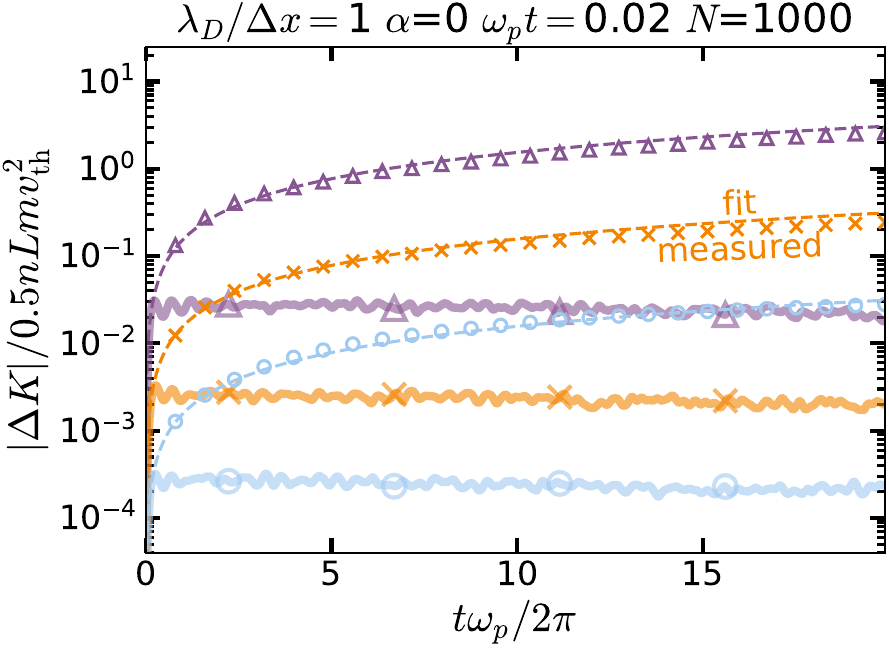}%
\raisesubfiglabel{0.05in}{2.1in}{(b)}%
\hspace{0.04in}%
\includegraphics*[width=2.1in]{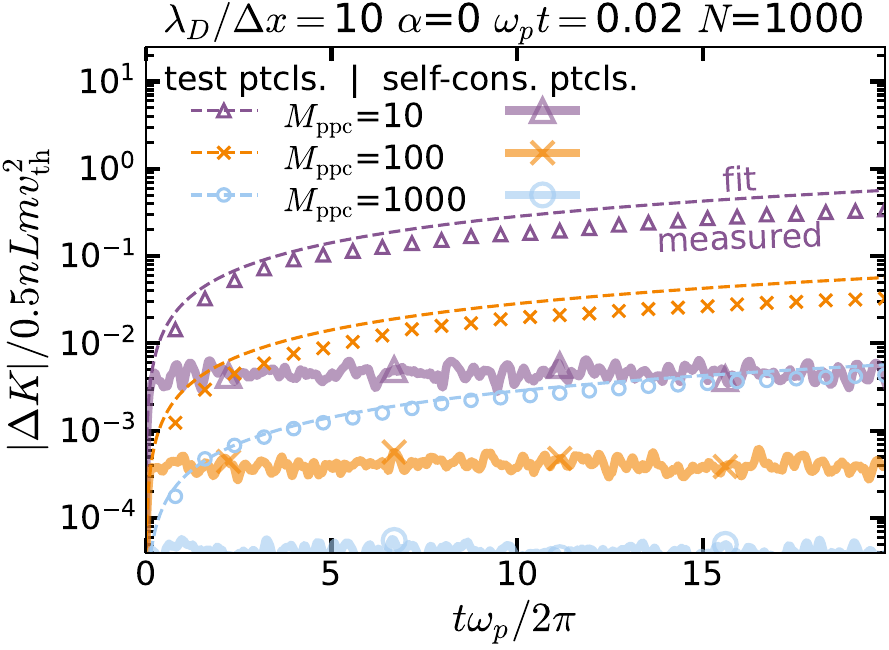}%
\raisesubfiglabel{0.05in}{2.1in}{(c)}%
\caption{\label{fig:testParticles}
The change in MC-PIC particle energy over time 
(normalized to the initial energy)
of self-consistent particles (thick, solid lines marked with occasional large symbols) compared with
test particles, over the first 20 plasma periods
for (from left to right) $\lambda_D/\Delta x=0.1$, $1$, and $10$.
For test particles, (relatively numerous) small symbols
show measurements, while thin, dashed lines show a simple fit using
Eq.~(\ref{eq:randWalk}) with $E_{\rm diff}=E_{\rm rms}$ from
Eq.~(\ref{eq:Erms}) and $\tau=\omega_p^{-1}$.
All 
simulations have $v_B=0$, $\omega_p \Delta t=0.02$, and no smoothing.
In each case, we show simulations with $M_{\rm ppc}=10$, $100$, and $1000$
(in purple, orange, and light blue).
}
\end{figure*}

In the previous subsection, we suggested that PIC particles 
exhibit stochastic velocity diffusion, but it is inconsistent with
experiencing a different random field of 
order $E_{\rm rms}$ 
every time interval~$\tau\sim \omega_p^{-1}$.
However, test particles---which do not interact---do exhibit this behavior.

Figure~\ref{fig:testParticles} shows $|\Delta K(t)|$ for the same
simulations as in Fig.~\ref{fig:quietVsNoisyStart} 
(with initially uniform 
positions and random velocities),
but for 20 plasma periods and for both 
self-consistent particles (thick, solid lines with infrequent, large symbols)
and test particles in the same simulation (more numerous, small symbols), 
for $M_{\rm ppc}=10$, $100$, and $1000$.  
The thin, dashed lines show a simple but surprisingly good fit to the test
particles,
the result of a random walk 
given by Eq.~(\ref{eq:randWalk}) with $E_{\rm diff}=E_{\rm rms}$ from
Eq.~(\ref{eq:Erms}) 
and time interval~$\tau=\omega_p^{-1}$.
The test particles always gain more energy than PIC particles.

For $\lambda_D/\Delta x=0.1$ the growth in self-consistent particle
energy is already significant, and the test particle energy
growth is worse, although not much worse in order of magnitude. 
For $\lambda_D/\Delta x=1$ and $M_{\rm ppc}=10$, 
the test particles double their energy over 20 plasma periods, 
compared with less than 2\% increase for self-consistent
particles; with $M_{\rm ppc}=1000$, test particles gain only about 3\%,
compared with 0.02\%. 
For $\lambda_D/\Delta x=10$
the use of test particles is more reasonable because $E_{\rm rms}$ is
relatively small; however, 
with $M_{\rm ppc}=10$ the test particle energy still grows
unphysically by 30\% over just 20 plasma periods, compared
with a fraction of a percent for
self-consistent particles.

Thus it is potentially misleading to use test particles 
as a proxy for self-consistent particles.

\section{Discussion}
\label{sec:discussion}

We have demonstrated that smoothing the charge density 
eliminates grid instability in explicit electrostatic MC-PIC simulation,
even when $\lambda_D \ll \Delta x$;
the key is that, for the smoothing profile given by Eq.~(\ref{eq:rhoSmoothing}),
the smoothing length $r_{\rm sm}$ must increase 
as $r_{\rm sm}/\Delta x \sim \Delta x/\lambda_D$.
Perhaps more important, we have shown that statistical particle noise
causes diffusion in velocity space, 
hence thermal energy that grows linearly in time, 
that worsens for smaller
$\lambda_D/\Delta x$, and at least in MC-PIC,
this poses a more pressing problem than grid instability.
Fortunately, smoothing suppresses noise as well as grid instability.
Unfortunately, smoothing degrades the effective spatial resolution
to (of order) the smoothing length~$r_{\rm sm}$.
While the preceding presentation focused on stabilizing
simulations 
with a given $\Delta x$, we now consider how to choose
$\Delta x$ and $r_{\rm sm}$ optimally, given a fixed resolution 
requirement.
What is the potential computational advantage of this approach?

A good estimate will be specific to the physics applicable to 
the particular problem being solved.  Here, 
we will assume that the problem is such 
the smallest important length scale is 
the scale on which the electric field and particle distribution vary
(as opposed to some other physics, such as a mean free path; e.g.,
some PIC simulations with Monte Carlo collisions 
might have their resolution determined by a mean free
path, even though the electric field and particle distribution vary
on much larger scales, and smoothing might not degrade the resolution for
collisions at all, leading to very different considerations).
We will write this smallest scale to be resolved as 
$\Delta x_{\rm ideal} \equiv \eta \lambda_D$,
where $\eta > 1$.

Under these conditions, the optimal simulation (with largest cell size)
will have
$r_{\rm sm}\sim \eta \lambda_D$ (since $r_{\rm sm}$ is the effective
resolution), 
hence $\Delta x \sim \eta^{1/2} \lambda_D$
(since $r_{\rm sm}/\Delta x \sim \Delta x/\lambda_D$).
I.e., the optimal cell size is the geometric mean of $\lambda_D$
and the required spatial resolution.

Compared with an unsmoothed (standard) simulation with cell size 
$\Delta x_s \sim \lambda_D$,
the increase in cell size $\Delta x/\Delta x_s \sim \eta^{1/2}$ can 
potentially speed up the field solve, presumably by at least a factor
$\eta^{1/2}$, but possibly $\eta^{p/2}$ for some power $p\geq 1$ that
depends on the field solve algorithm (and on the dimensionality, when
we consider 2D and 3D).
However, the smoothing operation is essentially a second field
solve, yielding a net gain of $\eta^{p/2}/2$,
if we ignore particle-related costs.

However, many PIC simulations are dominated by particle-related costs.
In such cases, reducing the field solve time does not speed up overall
computation.
Any gains require a reduction in the total number of particles.  
Since smaller $\lambda_D/\Delta x$ exacerbates noise, requiring more
particles per cell, suppressing noise is crucial.
With smoothing, $M_{\rm ppc}$ may be kept constant (at a small number, 
perhaps as small as 1 or 10) as $\Delta x$ is increased.
In that case, computation would speed up by a factor $\eta^{1/2}$ 
(and possibly by $\eta^{d/2}$ in $d$ dimensions, 
although PIC noise may scale differently in 2D and 3D,
so this requires future study).

In addition, if the simulation does not need to resolve the plasma
frequency well, the time step can be increased from roughly
$\Delta t \sim 
\Delta x/3 v_{\rm th} \sim \lambda_D/3v_{\rm th} = 1/3\omega_p$
(required to keep most particles from traveling more than
$\Delta x$ in time step $\Delta t$, when $\Delta x \sim \lambda_D$)
to $\Delta t \sim 2/\omega_p$ (for $\Delta x \gg \lambda_D$, 
the Courant-Friedrichs-Lewy 
time step is the limitation, not the maximum distance traveled).
Thus, in such cases,  larger cells and smoothing allow the time step
to be increased by~$\sim 6$.

A hypothetical ideal PIC algorithm might allow
larger cell size 
$\Delta x \sim \eta \lambda_D$.  
Although past attempts to
find such algorithms have focused on defeating grid instability, we
argue that particle noise is an equally urgent problem, if not more urgent.
Equation~(\ref{eq:dKthdt}) characterizes 
how noise-driven heating depends on numerical parameters $\Delta x$,
$M_{\rm ppc}$, and $r_{\rm sm}$.
Although practical simulations with non-uniform plasma may have other
considerations, this result for uniform plasma provides a good starting
point to determine optimal parameters for a given noise tolerance. 
For standard unsmoothed PIC ($r_{\rm sm}=0$),
the rate of noise-driven heating, $dK_{\rm th}/dt$, scales as 
$(\Delta x/\lambda_D)^3 M_{\rm ppc}^{-1}$.  If the scaling of the
PIC algorithm is known with respect to $\Delta x$ and $M_{\rm ppc}$,
we can find their optimal values for the maximum
tolerable~$dK_{\rm th}/dt$.
We suspect that most simulations will be optimized by using smaller
$\Delta x \sim \lambda_D$ rather than suffer the higher~$M_{\rm ppc}$ required to reduce noise when $\Delta x \gg \lambda_D$, even
if grid instability were not a consideration.
The introduction of smoothing ($r_{\rm sm} > \Delta x$) can keep
noise tolerable with larger $\Delta x$ and lower~$M_{\rm ppc}$
(as we demonstrated in the previous section),
since $dK_{\rm th}/dt$ scales as
$(\Delta x/\lambda_D)^3 [M_{\rm ppc} (1+(r_{\rm sm}/\Delta x)^2)
(1+r_{\rm sm}/\lambda_D)] ^{-1}$.
Ultimately, the optimal parameters will
depend on the PIC algorithm scaling, but it is likely that smoothing
can significantly reduce computational costs when under-resolving
the Debye length.

A different smoothing kernel might yield a more efficient
relationship between resolution and stability.
We leave such exploration to future work, but we reiterate that
the relationship $r_{\rm sm}/\Delta x \sim \Delta x/\lambda_D$ was
derived to stabilize the shortest (Nyquist) wavenumber
specifically for a smoothing kernel with Fourier transform
$(1+k^2 r_{\rm sm}^2)^{-1}\sim (kr_{\rm sm})^{-2}$.
In analogy with the derivation of Eq.~(\ref{eq:stabilityCriterion}), 
a smoothing kernel that, at $kr_{\rm sm} \gg 1$, scales as
$(k r_{\rm sm})^{-p}$ (for some $p>0$), 
will stabilize the Nyquist wavenumber for
$r_{\rm sm}/\Delta x \sim (\Delta x/\lambda_D)^{2/p}$.
Higher values of $p$ thus stabilize the shortest wavelengths with 
smaller~$r_{\rm sm}$, but stabilizing longer wavelengths might require
larger~$r_{\rm sm}$.

\section{Summary}
\label{sec:summary}

This paper focuses on three aspects of enabling electrostatic PIC
simulation with under-resolved Debye length: smoothing to defeat grid
instability, smoothing to defeat particle noise, and an efficient, 
flexible smoothing algorithm:
\begin{enumerate}[leftmargin=*]
  \item By writing down the MC-PIC dispersion 
  $D(\omega,k)$ with spatial discretization, we showed that
  for any fixed $k$, smoothing the charge density on the grid
  changes the dispersion in the same way as
  decreasing the plasma density (hence increasing its
  Debye length $\lambda_D$).  
  Knowing the change in density (or in $\lambda_D$) that would
  yield stability, we can calculate the smoothing that yields stability
  (for a given~$k$).
  This applies to any PIC algorithm with a dispersion
  of similar general form. 
   \begin{itemize}
   \item With the rough form of $D(\omega,k)$ and 
     empirical knowledge of the minimum stable density 
     (i.e., such that $\lambda_D \gtrsim \Delta x$),
     we determined the amount of smoothing needed to
     stabilize the shortest-wavelength modes. [\S\ref{sec:dispersion}]
   \item We find that smoothing with Eq.~(\ref{eq:rhoSmoothing})
     eliminates grid instability when
     the smoothing radius $r_{\rm sm}$ scales as
     $r_{\rm sm}/\Delta x =(\alpha/\pi) \lambda_D/\Delta x$, where
     $\alpha$ is a constant of order unity. [\S\ref{sec:dispersionResults}]
   \item Other smoothing profiles could potentially
     yield stability with smaller $r_{\rm sm}$. [\S\ref{sec:dispersion}]
   \item Without smoothing, the most unstable mode usually has
   $k\sim \pi/\Delta x$.
     Smoothing decreases the fastest-growing mode~$k$.
  [\S\ref{sec:dispersionResults}]
     \item
     By exhaustively finding the roots of $D(k,\omega)$ for
     a Cauchy-squared distribution, for many $k$,
     temperatures, and drift velocities,
     we found that, for the smoothing given by Eq.~(\ref{eq:rhoSmoothing}),
     the amount of smoothing 
     necessary for MC-PIC stability 
     is $\alpha\approx 5$ (although $\alpha\approx 2$
     might suppress growth enough for some applications). 
     [\S\ref{sec:dispersionResults}]
     \item We did not detect grid instability in MC-PIC simulations 
       (with Maxwellian distributions)
       with smoothing with $\alpha \geq 0.2$.  
       However, noise-driven heating was so strong
       for $\alpha \leq 1$ that grid instability might have been present
       while undetected. [\S\ref{subsec:suppression}]
   \end{itemize}

  \item We demonstrated how MC-PIC particle noise becomes increasingly
    disruptive as $\lambda_D/\Delta x$ decreases; 
    for simulations with $\lambda_D < \Delta x$,
    noise may be an equally or even more important consideration than
    grid instability.
    \begin{itemize}
     \item  
       Noise-driven heating grows linearly in time, whereas
       grid-instability heating grows exponentially
        [\S\ref{subsec:signatures}].
      \item 
        Noise-driven heating rates are proportional to~$M_{\rm ppc}^{-1}$,
        whereas grid instability growth rates
        become independent of $M_{\rm ppc}$ at
        large $M_{\rm ppc}$.
        [\S\ref{subsec:signatures}]
      \item
        E.g., for $\Delta x \approx 5 \lambda_D$, grid instability 
        is undetectable even with $M_{\rm ppc}=1000$, while
        noise-driven heating is ruinous.
        [\S\ref{sec:dispersionResults}, \S\ref{sec:PICresults}]
      \item 
        For $\Delta x \gtrsim \lambda_D$, it is likely more 
        computationally efficient
        to reduce noise by reducing $\Delta x$ (keeping the total
        number of particles fixed) than by increasing $M_{\rm ppc}$.
        [\S\ref{subsec:velocityDiffusion}]
      \item The amount of smoothing
         necessary to eliminate grid instability is roughly the same as
         the amount needed to suppress noise-driven heating in MC-PIC
         for small $M_{\rm ppc}$.
         [\S\ref{sec:dispersionResults}, \S\ref{subsec:suppression}]
    \item The PIC ``noise electric field'' on the cell scale in 1D,
    due to \emph{uncorrelated} random particle positions
    is given by $E_{f\Delta x}$ in Eq.~(\ref{eq:randomNoise}) with
    $f=M_{\rm ppc}^{-1}$.
      The energy density of this field is
  $\epsilon_0 E_{f\Delta x}^2/2 \sim 
     M_{\rm ppc}^{-1} n m v_{\rm th}^2 (\Delta x/ \lambda_D)^2/2$.
    For $\Delta x \gg \lambda_D$, this unphysical 
    field energy can easily exceed the plasma energy.
      This may be a general concern for any PIC simulation in
      which particles obtain positions due to mechanisms other than the
      fundamental PIC algorithm (e.g., initial loading, injection from
      boundaries, collisions, combining macroparticles).
    [\S\ref{sec:noiseTheory}, \S\ref{subsec:noisyStart}]
    \item However, the self-consistent electric 
      field has smaller fluctuations than $E_{f\Delta x}$
      because particle positions are correlated even on sub-grid
      scales.
      We find empirically in MC-PIC
      that the root mean square field in 
      uniform plasma is
\begin{eqnarray}
  E_{\rm rms} &\sim & \frac{m v_{\rm th}^2}{e \lambda_D}
  \sqrt{\frac{1}{(1+\lambda_D/\Delta x)(1+r_{\rm sm}/\Delta x)}}
    M_{\rm ppc}^{-1/2}
\end{eqnarray}
    where $r_{\rm sm}$ is the radius of smoothing used.
    This agrees roughly with analytical calculations of 
    \citet{Touati_etal-2022}, and adds the correction for $r_{\rm sm}>0$.
    [\S\ref{subsec:Enoise}]
    \item Initialization with random particle positions 
      is disastrous for $\lambda_D \lesssim \Delta x$, unless
      $M_{\rm ppc}$ is much higher than typically used in PIC.
    [\S\ref{sec:noiseTheory}, \S\ref{subsec:noisyStart}]
    \item A 1D MC-PIC simulation of uniform plasma 
      (initialized with uniformly-spaced particles with
      random velocities)
      exhibits three
    stages of energy change (or two stages, without smoothing).
        [\S\ref{subsec:velocityDiffusion}]
    \subitem
    (1) Almost immediately, the electric field grows to $E_{\rm rms}$
     at the expense of plasma thermal energy,
    roughly conserving energy overall. 
    The expression for $E_{\rm rms}$ shows that
    for $\lambda_D \lesssim \Delta x$ and $r_{\rm sm}=0$, 
      this is a relaxation to numerical equipartition (i.e., between electric field and macroparticle degrees of freedom); 
        smoothing reduces this unphysical energy exchange.
    \subitem
    (2) With smoothing, $r_{\rm sm} \gtrsim \Delta x$, the plasma 
      thermal energy continues to decline by a 
      fraction~$\sim 0.3/M_{\rm ppc}$
      from its initial value over a time
      $t_{\rm decl} \sim (1+r_{\rm sm}/\Delta x)/(\sqrt{2}\omega_p)$.
      The electric field energy does not increase correspondingly,
      violating energy conservation.
      The total decline is independent of $r_{\rm sm}$,
      although smoothing helps by increasing $t_{\rm decl}$.
    \subitem
    (3) The plasma energy grows linearly in time due to stochastic
    velocity diffusion (noise-driven heating).  
    The electric field energy grows correspondingly, violating
    energy conservation.
    \item We find that, in 1D MC-PIC, these three stages of changing thermal
      energy $K_{\rm th}$ can be described quantitatively by:
\begin{widetext}
\begin{eqnarray} 
  \frac{1}{\omega_p K_{\rm th}(t)}
  \frac{dK_{\rm th}}{dt}(t)
  &\sim &
  \left\{ \begin{array}{l@{\quad\textrm{for }}l}
    -\frac{\delta(\omega_p t)}{2 M_{\rm ppc} 
      (1+\lambda_D/\Delta x)(1+r_{\rm sm}/\Delta x)}
      & \omega_p t \approx 0 \\
    -\frac{0.3 \sqrt{2}}{M_{\rm ppc}(1+r_{\rm sm}/\Delta x)}
    \phantom{\frac{\displaystyle 1^1}{\displaystyle 1^1}}
      & 0 \lesssim \omega_p t \lesssim 
                  \frac{1+r_{\rm sm}/\Delta x}{ \sqrt{2}}
                  \\
  \frac{ 5 \times 10^{-4}  
    \left( \frac{\Delta x}{\lambda_D(t)} \right)^3 }{
    M_{\rm ppc}
    \left[ 1 + \left(\frac{r_{\rm sm}}{\Delta x}\right)^2 \right]
    \left( 1 + \frac{r_{\rm sm}}{\lambda_D(t)} \right)
  }
           & \frac{1+r_{\rm sm}/\Delta x}{ \sqrt{2}} 
                 \lesssim  \omega_p t 
  \end{array} \right.
\end{eqnarray}
\end{widetext}
  Noise-driven heating (the bottom line) increases strongly
  as $\lambda_D/\Delta x$ decreases below~1.
    [\S\ref{subsec:velocityDiffusion}]

    \item Test particles
      experience greater velocity diffusion 
      than self-consistent particles.
      In MC-PIC, test particles diffuse as if receiving a random kick of
      strength $E_{\rm rms}$ every time interval $\tau = \omega_p^{-1}$.
      Using test particles (rather than tracking a sample of
      self-consistent PIC particles) may yield misleading 
      results.
      [\S\ref{subsec:testParticles}]
    \end{itemize}
  \item Severely under-resolving
  the Debye length requires smoothing over many cells;
  finding a performant smoothing algorithm may be nontrivial.
  We proposed and tested a smoothing
  algorithm involving
  a modified Poisson solve, which is very similar
  to the Poisson solve used to find the electric potential, except more
  diagonally dominant.  With this approach, we hope the penalty 
  for smoothing will less than the time required for the
  already-existing Poisson solve for the potential.
  \begin{itemize}
  \item The modified Poisson-smoothing offers similar flexibility
  as the potential solve.  
  For example, if a PIC code has a Poisson solver that
  can handle variable cell sizes, it can
  likely be easily modified to perform the smoothing operation for
  the same conditions.
      [\S\ref{sec:poissonSmoothing}]
  \end{itemize}
\end{enumerate}

\vspace{\baselineskip}

To confirm the effectiveness of smoothing, we showed stable 
MC-PIC simulation
(of a uniform, drifting 1D plasma) with tiny
$\lambda_D/\Delta x = 1/2000$ while using just 1 macroparticle
per cell, for $10^4$ plasma periods
      (cf.~\S\ref{subsec:suppression}).

Smoothing degrades the effective spatial resolution of the electric field.
For the smoothing described in~\S\ref{sec:poissonSmoothing},
the optimal cell size (to achieve an effective electric field resolution 
of $r_{\rm sm}$) is $\Delta x \sim \sqrt{\lambda_D r_{\rm sm}}$
(cf.~\S\ref{sec:discussion}).

Although we demonstrated this method in standard 1D electrostatic, 
explicit momentum-conserving PIC
for regular Cartesian mesh 
and periodic boundary conditions, we expect all the elements 
(including smoothing via modified Poisson solve) to generalize 
to 2D and 3D, other PIC variants, 
non-Cartesian and possibly irregular meshes, 
with embedded metal boundaries.
In particular the smoothing via Poisson solve allows 
spatially-varying~$r_{\rm sm}$, 
which could be used with variable meshes to resolve 
sheaths near boundaries or regions of locally dense plasma,
while under-resolving the Debye length in the bulk plasma.

Smoothing the current density might well suppress grid instability and 
noise in electromagnetic PIC algorithms.  However, explicit 
electromagnetic codes likely
have a field advance significantly faster than the 
modified Poisson solve needed for smoothing, so smoothing may be
relatively more expensive than in electrostatic PIC.

Although we demonstrated the effectiveness of smoothing with a particular
smoothing operation, other smoothing methods and
profiles might work as well or better.
Developing optimal smoothing methods---in terms of speed and performing
the minimum necessary smoothing---could potentially provide important benefits.

Since we considered only explicit PIC algorithms, we did not explore
potential gains from increasing the time step beyond $2/\omega_p$.
An efficient algorithm that allows this might be combined with smoothing
to great advantage (for applications that do not need to resolve the
plasma frequency, i.e., in which even long-wavelength Langmuir modes are 
unimportant).  Moreover, because such implicit algorithms typically
decrease the simulation plasma frequency below $2/\Delta t$, 
its imaginary component may be similarly decreased, 
slowing the grid instability
\citep{Brackbill_Forslund-1982}.  However, such simulations might
still require substantial smoothing to reduce particle noise.

The potentially worst problem we observed with the
proposed method
for under-resolving the Debye length was the slow unphysical decline of
initial plasma thermal energy by a fraction~$\sim 0.3/M_{\rm ppc}$, unaffected
by smoothing.  That may limit feasibility for simulations with
$\Delta x \gg \lambda_D$ if it means that they require $M_{\rm ppc} \gg 1$.
While $M_{\rm ppc} \gg 1$ is reasonable for PIC simulations where
$\Delta x$ is set to the desired resolution (i.e., the scale on which
the plasma varies significantly), it may be an onerous requirement
when
$\Delta x$ has to be much smaller than the effective resolution,
$r_{\rm sm}$.
The fundamental problem here is perhaps one of equipartition: the need
for smoothing to suppress sub-grid modes results in $\Delta x$ being
smaller than ideal, leaving the
electric field with more degrees of freedom than desirable.
Future work may need to address this problem; however, at least this
is not a continuing instability.

Besides investigating smoothing in 2D and 3D, further 
work might also explore whether the smoothing length could be
set dynamically based on the local plasma density to ensure stability.
An especially promising aspect of this method is that it may allow
high resolution to resolve $\lambda_D$ near boundaries for accuracy
in modeling sheaths, while under-resolving $\lambda_D$ in the bulk
plasma.

\section*{Acknowledgments}

We would like to thank Thomas G. Jenkins and Daniel Main for many helpful discussions.
This work was supported by the National Science Foundation, 
grants PHY 2206904 and PHY 2206647.

\section*{Author declarations}

The authors have no conflicts to disclose.

\section*{Data availability}

The data that support the findings of this study (including a Python script to run PIC simulations and a Mathematica script to find roots of the dispersion) are openly available in the Zenodo repository at 
\href{https://doi.org/10.5281/zenodo.14984906}{doi:doi:10.5281/zenodo.14984906}.

\appendix 

\section{Deriving the finite difference dispersion}
\label{sec:dispersionDerivation}

The finite-difference plasma dispersion has been derived previously 
\citep{Langdon-1970,BirdsallPIC}.  We show it here
for completeness and consistency of notation (e.g., with MKS units).

We want to find the Landau damping (or inverse Landau growth rates) for
the 1D Vlasov-Poisson system in a uniform electron plasma with
a uniform immobile ion background:
\begin{eqnarray}
  \partial_t f(x,v) &=& -v \partial_x f(x,v) -  a(x) \partial_v f(x,v) \\
  a(x) &=& -\frac{e}{m} E(x) = \frac{e}{m} \partial_x \phi(x) \\
    -\partial_x^2 \phi(x) &=& [\rho_{\rm ion} + \rho(x)]/\epsilon_0
\end{eqnarray}
where $-e$ and $m$ are the electron charge and mass, 
$\phi(x)$ the electric potential, and $f(x,v)$ the electron distribution;
$a(x)$ is the acceleration of an electron at~$x$ due to electric field
$E(x)$.  The electron charge density is $\rho(x)=-e\int f(x,v)dv$, and
$\rho_{\rm ion} = -L^{-1}\int_0^L \rho(x) dx$ 
is the uniform ion charge density.

It suffices for this paper (and follows earlier treatments) 
to discretize the Vlasov-Poisson system
in infinite space, but not in time, nor in particles---this 
will correspond to PIC in the 
limit of an infinite number of simulated particles with tiny
time step.

The simplest discretization is the field solve, which involves
only the grid with nodes $x_n=n \Delta x$ for 
integers~$n$.

We use a standard 3-point finite difference (FD) Poisson stencil
to relate the gridded charge density $\rho_n$ and $\phi_n$:
\begin{eqnarray}
  (-\nabla_{FD}^2 \phi)_n &\equiv & 
  \frac{-\phi_{n+1} + 2 \phi_n - \phi_{n-1}}{\Delta x^2} 
  = \frac{\rho_{\rm ion}+\rho_n}{\epsilon_0}
\end{eqnarray}
where $\phi_n\equiv \phi(x_n)$ and $\rho_n\equiv \rho(x_n)$.
We then solve for the edge-center electric field $E^e_{n+1/2}$, and then
the nodal field
\begin{eqnarray}
  E^e_{n+1/2} &=& - (\phi_{n+1} - \phi_n)/\Delta x \\
  E_n &=& (E^e_{n-1/2} + E^e_{n+1/2})/2
.\end{eqnarray}

The trickier aspect of
discretization involves depositing a particle's charge to the
grid charge density, $\rho_n$, and interpolating the grid (nodal
electric field) $E_n$ to particle positions
To to this, 
we introduce the shape function $S(x_m;x) = S(x_m-x)$ which describes
(1) the fraction of charge of a particle at position $x$ that should be
deposited to the grid node $x_m$, and (2) the weight of the nodal
electric field $E_m\equiv E(x_m)$ when interpolating the field
to particle position~$x$.
In principle these shape functions can be different, but standard 
momentum-conserving PIC codes use the same shape for both, and using
different functions would not affect the following analysis.
With this, we have:
\begin{eqnarray}
  \rho_n &=& \frac{1}{\Delta x} 
    (-e) \int dv \int dx f(x,v) S(n\Delta x - x) \\
  a(x) &=& -\frac{e}{m} E(x) = -\frac{e}{m} \sum_n E_n S(n\Delta x - x)
.\end{eqnarray}
After linearization about a plasma with uniform distribution
$f_0(x,v) = n_0 F_0(v)$ (where $n_0$ is the plasma density), 
hence zero electric field, 
the spatially-discretized equations are, treating
$f(x,v)$, $\phi_n$, $E_n$, and $\rho_n$ 
as first-order perturbed quantities:
\begin{widetext}
\begin{eqnarray}
  \partial_t f(x,v) + v \partial_x f(x,v)
  &=&
  \frac{en_0}{m} \sum_n E_n S(n\Delta x - x) \partial_v F_0(v) 
  \\
  (-\nabla_{FD}^2 \phi)_n &= & \frac{\rho_n}{\epsilon_0} =
  -\frac{e}{\epsilon_0 \Delta x}
    \int dv \int dx f(x,v) S(n\Delta x - x) \\
  E^e_{n+1/2} &=& - (\phi_{n+1} - \phi_n)/\Delta x \\
  E_n &=& (E^e_{n-1/2} + E^e_{n+1/2})/2
\end{eqnarray}
\end{widetext}
We have not spatially discretized $f(x,v)$ or 
its advective derivative 
$v \partial_x f$, because we will evolve $f(x,v)$ via the method
of characteristics, represented as the
sum of discrete particles; i.e., we will not
be evaluating $\partial_x f$ directly.

We Fourier transform in time and space, 
assuming ``modes'' $e^{i(kx-\omega t)}$ (which are not 
eigenmodes, because the grid breaks translational symmetry),
using the following conventions for gridded 
(e.g., $E_n$) and continuous (e.g., $S(x)$) quantities
\begin{eqnarray}
  \tilde{E}_k = \sum_n e^{-ikn\Delta x} E_n 
  &\qquad &
  E_n = \frac{\Delta x}{2\pi} \int_{-\pi/\Delta x}^{\pi/\Delta x} 
             e^{ikn \Delta x} \tilde{E}_k dk
  \\
  \tilde{S}_{q} = \frac{1}{\Delta x}
       \int_{-\infty}^{\infty} e^{-iqx} S(x) dx 
     &&
  S(x) = \frac{\Delta x}{2\pi} 
           \int_{-\infty}^{\infty} e^{iqx} \tilde{S}_q dq
\end{eqnarray}
% [
where $k \in (-\pi/\Delta x,\pi/ \Delta x]$,  %)
and $q$ is any real value, which we will
often write, uniquely, as $q\equiv k + k_g$ for integer $g$, 
where $k_g \equiv 2\pi g/\Delta x$ (hence $e^{ik_g n \Delta x}=1$).
The following transforms of convolutions/products will be useful:
\begin{eqnarray}
    \frac{1}{\Delta x} \int e^{-i(k+k_g) x}
    \sum_n E_n S(n\Delta x - x) dx
    &=&
    \tilde{E}_{k} \tilde{S}_{-k-k_{g}} 
    \\
    \sum_n e^{-ikn\Delta x} \int S(x_n-x) f(x) dx
    &=&
      \Delta x \sum_g \tilde{f}_{k+k_g} \tilde{S}_{k+k_g}
\end{eqnarray}

In Fourier space, the gridded field solve can be described as
\begin{eqnarray}
  K(k)^2 \tilde{\phi}_k &=&  \tilde{\rho}_k / \epsilon_0 \\
 -i \kappa(k) \tilde{\phi}_k &=& \tilde{E}_k
,\end{eqnarray}
where, in continuous space we would have $K(k)^2=k^2$ and $\kappa(k) = k$,
but 
with our discretization, 
$K(k)^2 = (4/\Delta x)^2 \sin^2(k\Delta x/2)$ and
$\kappa(k) = \sin(k\Delta x)/\Delta x$.

The linearized Vlasov-Poisson equations become
\begin{eqnarray}
  -i[\omega - (k+k_g) v] 
  \tilde{f}_{k+k_g}
  &=& 
  \frac{en_0}{m}  
      \tilde{E}_{k} \tilde{S}_{-k-k_{g}} 
      \partial_v F_0(v)
    \label{eq:lVP1} \\
    \tilde{E}_k &=& 
    -i \kappa(k) \tilde{\phi}_k
    =
    -i \frac{\kappa(k)}{\epsilon_0 K(k)^2} \tilde{\rho}_k
    \label{eq:lVP2} \\
  &=&
   i\frac{e}{\epsilon_0 } 
     \frac{\kappa(k)}{K(k)^2}
      \sum_g \tilde{S}_{k+k_g} \int \tilde{f}_{k+k_g}(v) dv
      \nonumber 
\end{eqnarray}
Dividing Eq.~(\ref{eq:lVP1}) by $-i[\omega-(k+k_g)v]$ and integrating
over $v$, and substituting into Eq.~(\ref{eq:lVP2}),
\begin{widetext}
\begin{eqnarray}
  \int \tilde{f}_{k+k_g} dv
  &=& 
  \frac{en_0}{m}  \tilde{E}_{k} \tilde{S}_{-k-k_{g}} 
      \int \frac{\partial_v F_0}{ -i[\omega - (k+k_g) v] } dv
    \\
    \tilde{E}_k &=& 
   i\frac{e^2 n_0}{\epsilon_0 m } 
     \frac{\kappa(k)}{K(k)^2}
      \sum_g \tilde{S}_{k+k_g} 
      \tilde{E}_{k} \tilde{S}_{-k-k_{g}} 
      \int \frac{\partial_v F_0}{ -i[\omega - (k+k_g) v] } dv
      \nonumber \\
  &=&
   -\omega_p^2 
   \left[
     \frac{\kappa(k)}{K(k)^2}
      \sum_g |\tilde{S}_{k+k_g}|^2
      \int \frac{\partial_v F_0}{ \omega - (k+k_g) v } dv
      \right]
      \tilde{E}_{k} 
\end{eqnarray}
\end{widetext}
Thus we write the dispersion 
\citep[compare to Sec.~8-10 of][]{BirdsallPIC}
\begin{eqnarray} 
  D(\omega,k) &=& 1 + 
   \frac{\omega_p^2 }{K(k)^2}
     \kappa(k)
      \sum_g |\tilde{S}_{k+k_g}|^2
      \int \frac{\partial_v F_0}{ \omega - (k+k_g) v } dv
      \nonumber \\
\label{eq:dispersionAppendix}
      &=&
      1 -
   \frac{\omega_p^2 }{K(k)^2}
     \kappa(k)
      \sum_g \frac{|\tilde{S}_{k+k_g}|^2}{k+k_g}
      \int \frac{\partial_v F_0}{ v - \omega / (k+k_g) } dv
\end{eqnarray}

\section{The dispersion integral for the Cauchy-squared distribution} 
\label{sec:dispersionIntegral}

For the Cauchy-squared distribution, Eq.~(\ref{eq:CauchySqrDist}), 
the integral in the dispersion, 
Eq.~(\ref{eq:dispersion}) or Eq.~(\ref{eq:dispersionAppendix}), can be written
%\begin{widetext}
\begin{eqnarray}
  \int \frac{\partial_v F_0(v) \; dv}{v - \omega / (k+k_g)} 
  &=&
  \int \frac{F_0(v) \; dv }{[v - \omega / (k+k_g)]^2} 
  \\
  &=&
  \int \frac{ (2 v_{\rm th}^3 / \pi) \, F_0(v) \; dv}{
      \left[v_{\rm th}^2+(v-v_B)^2 \right]^2 
      \left[v - \omega / (k+k_g) \right]^2} 
      \nonumber 
\end{eqnarray}
%\end{widetext}
This integral should be calculated using the Landau prescription described
by many plasma textbooks \citep[e.g.,
the dispersion can be evaluated just as for a Cauchy distribution
described in detail in Ch.~13 of][]{Gurnett_Bhattacharjee-Book}.
For Im$[\omega]>0$, the Landau prescription uses 
straightforward contour integration (deviating around any poles).
This integral is a rational function of $\omega$, and therefore
the result it is identical to its analytic continuation to Im$[\omega]<0$.
Thus we can skip the rest of the Landau prescription, which is just a 
recipe for finding the analytic continuation, and conclude that
for all $\omega$,
\begin{eqnarray}
  \int \frac{\partial_v F_0(v) \; dv}{v - \omega / (k+k_g)} 
  &=&
  (k+k_g)^2 \,
  \frac{ \omega - (k+k_g)v_B + 3i |k+k_g|v_{\rm th} }{
      \left[ \,\omega - (k+k_g)v_B + i |k+k_g|v_{\rm th}\, \right]^3 }
.\end{eqnarray}

\section{The exact dispersion for a Cauchy-squared distribution}
\label{sec:dispersionExact}

Before solving the numerical dispersion in the next section,
we consider the exact physical dispersion for a Cauchy-squared
distribution, Eq.~(\ref{eq:CauchySqrDist}), in continuous space.
The continuous-space dispersion is given by 
Eq.~(\ref{eq:dispersionAppendix}),
but keeping only the term with $k_g=0$.
Using the integral from Appendix~\ref{sec:dispersionIntegral},
\begin{eqnarray} \label{eq:physicalDispersion}
  D_{c}(\omega,k) &=& 1 
  - \frac{\omega/\omega_p - kv_B/\omega_p + 3 i |k| \lambda_D}{
   \left[\omega/\omega_p - kv_B/\omega_p + i |k| \lambda_D \right]^3}
.\end{eqnarray}
Solving $D_c=0$ requires solving a cubic equation; its 3 solutions
are $\omega_0$ and $\omega_\pm$, shown below with their limits for 
$k\lambda_D \rightarrow 0$: 
\begin{widetext}
\begin{eqnarray}
  \frac{\omega_\pm}{\omega_p}
  &=&
  \frac{k v_B}{\omega_p} - i |k| \lambda_D
  +
  \frac{
    \pm ( b^{1/3} + b^{-1/3} )
    + i \, 3^{-1/2} ( b^{1/3} - b^{-1/3}) 
      }{2}
      \\
  && \rightarrow \quad
  \frac{k v_B}{\omega_p} \pm (1 + 3 k^2\lambda_D^2/2)
  - 4 i |k|^3 \lambda_D^3
    \qquad (\textrm{as } k\lambda_d \rightarrow 0)
  \nonumber \\
  \frac{\omega_0}{\omega_p}
  &=& 
  \frac{k v_B}{\omega_p} - i |k| \lambda_D
  - i \, 3^{-1/2} ( b^{1/3} - b^{-1/3}) 
  \quad \rightarrow \quad - 2 i |k| \lambda_D
                   \\
  && \textrm{where } b \equiv  
       \sqrt{27k^2\lambda_D^2} + \sqrt{1 + 27 k^2 \lambda_D^2} 
    \nonumber
\end{eqnarray}
\end{widetext}
The real part of $\omega_\pm$ yields the Bohm-Gross dispersion, exactly as
with a Maxwellian or with any one-humped distribution
with the same first and second velocity moments.
These Langmuir waves are weakly-damped for $|k|\lambda_D \ll 1$ and
strongly-damped for $|k|\lambda_D \gtrsim 1$, qualitatively similar to
a Maxwellian 
but differing in dependence on $k\lambda_D$ 
\citep{Gurnett_Bhattacharjee-Book}.
In addition, there is a third, purely-decaying mode~$\omega_0$; 
we will not be concerned with this harmless mode,
keeping in mind that the (initial-value) Landau approach does not yield 
true eigenmodes,
much less all the eigenmodes of a system with an infinite number
of degrees of freedom 
\citep[a topic well beyond the scope of this paper; see, e.g.,][]{VanKampen-1955,Case-1959}.

\section{Solving the numerical dispersion for a Cauchy-squared distribution}
\label{sec:dispersionSolving}

The numerical dispersion, 
Eq.~(\ref{eq:dispersion}) or Eq.~(\ref{eq:dispersionAppendix}), 
looks like the
exact physical dispersion, Eq.~(\ref{eq:physicalDispersion}), 
with more poles corresponding to aliases of $k$
(i.e., $k+k_g$); we can rewrite it as follows
\begin{eqnarray} \label{eq:CauchyDispersion}
  D &=& 1 - a_k \sum_g b_{k+k_g} 
    \frac{x - x_{k+k_g} + 2 i |y_{k+k_g}|}{(x-x_{k+k_g})^3}
    \\
   &=&  1 - a_k \sum_g b_{k+k_g} 
    \frac{1}{(x-x_{k+k_g})^2}
    \left[ 1 + \frac{2 i |y_{k+k_g}|}{x-x_{k+k_g}} \right]
    \nonumber
\end{eqnarray}
for normalized frequency $x\equiv \omega/\omega_p$, with poles
(for all integer $g$)
$x_{k+k_g}\equiv (k+k_g)v_B/\omega_p - i |y_{k+k_g}|$,
$y_{k+k_g}\equiv (k+k_g)\lambda_D$,
and $a_k$ and $b_{k+k_g}$ are related to the field solve and shape
functions.  Using a linear shape function (shaped like a tent spanning 
width $2\Delta x$):
\begin{eqnarray}
  S(n\Delta x-x) &=& 
    \Theta\left(\Delta x - |x-n\Delta x| \right ) \,
    \frac{|x - n\Delta x|}{\Delta x} \\
  \tilde{S}_{k+k_g} &=&  \frac{4\sin^2(k\Delta x/2)}{(k+k_g)^2 \Delta x^2}
\end{eqnarray}
(where $\Theta$ is the Heaviside step function),
we obtain $b_{k+k_g} = [(k+k_g)\Delta x]^{-3}$.
For a Fourier solve, with $K^2=k^2$ and $\kappa=k$, we get
$a_k = 16 \sin^4(k\Delta x/2) / (k \Delta x)$;
for the standard finite difference field solve we use
[see explanation of $K$ and $\kappa$ after Eq.~(\ref{eq:dispersion})],
$a_k = 4 \sin^2(k\Delta x/2) \sin(k\Delta x)$.

Our strategy to find dispersion roots approximates 
$D(k,x\omega_p)$ for $x$ within a circle centered at $x_c$ 
(eventually for
many different circles covering the range of interest).  Within the
circle, we retain some terms of Eq.~(\ref{eq:CauchyDispersion}) 
exactly---namely terms with poles within and near to the circle.
The remaining terms are Taylor-approximated as a polynomial in $(x-x_c)$
that is guaranteed to converge within the circle (because there are
no poles of $D$ there).
Importantly, we can place a rigorous upper bound on the error of
truncating the Taylor series, as well as on the error of
truncating the sum over $g$.
We can always increase accuracy, while keeping truncation orders the same,
by shrinking the size of the circle (and using more circles to
cover the area of interest).
By multiplying the approximation (of known accuracy) by all its poles, 
we obtain a polynomial in $x$, for which can reliably find all the roots.

In more detail,
the dispersion $D(k,\omega)$ in Eq.~(\ref{eq:CauchyDispersion}) is one
minus a sum of ``pole terms'': i.e., each term contains one
pole $x_{k+k_g}$ (of order 3).
To find roots of $D(k,\omega)$
reliably within a chosen area of the complex 
$\omega$ plane (or, equivalently, the complex $x$ plane) 
for one $k$ at a time, we cover that area with a set
of circles; in each circle we construct a polynomial in $x$ with
zeros that approximate the zeros of $D(k,x\omega_p)$; we then find 
all the polynomial's zeros, keeping those that lie in the circle.
To approximate $D$ around $x_c$ at the center of a circle (with radius
to be determined by the approximation's accuracy),
we keep $N_{\rm exact}$ pole terms exactly as they appear in
Eq.~(\ref{eq:CauchyDispersion}),
always including $g=0$, plus the $N_{\rm exact}-1$ poles nearest $x_c$.
We find $R$, the distance from $x_c$ to the 
first pole not among the $N_{\rm exact}$ poles.
For the rest of the terms, corresponding to poles outside a circle of
radius $R$ around $x_c$, we perform a Taylor expansion in $x$
around $x_c$.  
Because the poles of all Taylor-approximated terms lie outside $R$, 
the Taylor expansions converge for $|x-x_c|<R$.
Within a given radius $r<R$, it is tedious but straightforward to 
place a reasonable upper bound on the error accrued by truncating the
Taylor expansion to order $N_{\rm Taylor}$; the error is highest for the
$g=\pm 1$ terms, and using that as a conservative estimate for all 
$|g|>1$ simplifies
the process.  The smaller $r/R$ is, the smaller $N_{\rm Taylor}$ can
be (for a desired error).
We can also determine (for a given $r$) where we can 
safely truncate the sum over $g$ in Eq.~(\ref{eq:CauchyDispersion}),
i.e., choosing $G$ and keeping terms $|g|<G$) without exceeding the
desired error.
With a straightforward 1D numerical root solve, we can find the largest
$r$ that yields the desired accuracy, given 
$N_{\rm Taylor}$ and $N_{\rm exact}$.

With $N_{\rm exact}$ exact terms in the sum in 
Eq.~(\ref{eq:CauchyDispersion}), and the rest of the terms either
neglected ($|g|>G$) or Taylor-approximated by a polynomial of order
$N_{\rm Taylor}$, we can write $D$ as a rational function of $x$;
multiplying the equation
$D=0$ by the product of $(x-x_{k+k_g})^3$ for the $N_{\rm exact}$ poles,
we arrive at a polynomial equation of degree 
$3N_{\rm exact}+N_{\rm Taylor}$.  We find all the roots of this
polynomial, and then keep only the roots within radius $r$ of $x_c$.
We do this for all the circles $(x_c,r)$, which cover the entire area 
of interest, and collect all the roots.  Because the circles necessarily
overlap, some roots inevitably appear multiple times, with slight
differences below the maximum allowed error.

We use Mathematica to do this, and once we form the approximate 
polynomial (for a given circle), we find its roots using a numerical
precision of 256 digits
(for comparison, standard double precision has about 16 digits).  
This has been sufficient to find roots with an absolute error less than 
$10^{-6}$ for polynomials of degree up to about 64;
we ``spend'' about 32 degrees on the Taylor expansion, allowing inclusion
of about 10 poles exactly.

\begin{figure*}
\centering 
\includegraphics*[width=3.2in]{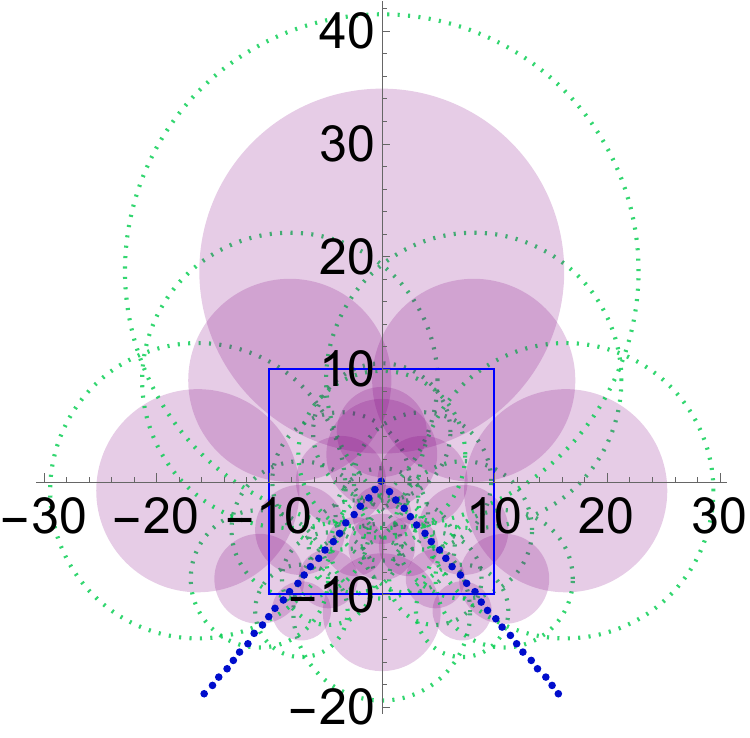}%
\raisesubfiglabel{-0.11in}{3.2in}{(a)}%
\includegraphics*[width=3.2in]{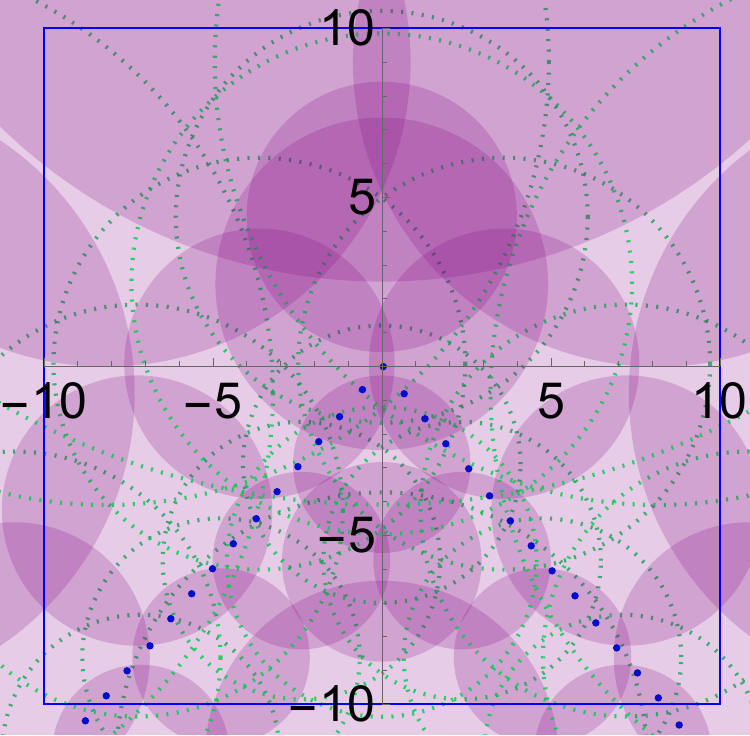}%
\raisesubfiglabel{-0.11in}{3.2in}{(b)}%
\caption{ \label{fig:covering}
The 22 circles in the complex $\omega$ plane covering
region of interest (blue box), 
$-10 \leq \textrm{Re}[\omega]/\omega_p \leq  10$ on the horizontal
axis and
$-10 \leq \textrm{Im}[\omega]/\omega_p \leq  10$ on the vertical, for
$v_B=0.1\omega_p \Delta x$ and $v_{\rm th}=0.12\omega_p \Delta x$
(hence $\lambda_D/\Delta x=0.12$).  
Panel (b) is the same as (a), but zoomed in.
The blue dots show the poles 
of $D(k,\omega)$.  For each magenta disk (of radius $r$) $D$ was
approximated by a different polynomial to find its roots; corresponding
to each magenta disk is a dotted green circle showing the radius $R$
containing no more than about 11 poles (the $N_{\rm exact}$ poles).
}
\end{figure*}

We will not describe the lengthy, systematic process for finding the
covering of circles, but show an
example in~Fig.~\ref{fig:covering}.  The transparent magenta disks
(radius~$r$) are
the regions $|x-x_c|<r$ in which the Taylor expansions 
are accurate to a part in $10^6$ or better; around each magenta disk is 
a dotted green circle (radius~$R>r$) that contains no more than 11 poles.
By including these terms exactly, we thus ensure that 
the Taylor expansions of the remaining terms converge everywhere
in $R$.
The Taylor expansions converge within
the dotted green circle (radius $R$), 
but are guaranteed to be sufficiently accurate only in the magenta.
The poles themselves
(blue dots) are located along the edges of a $\Lambda$-shaped wedge 
with vertex 
at the origin ($\omega=0$); a big advantage of using the Cauchy-squared
distribution is that the pole locations are easily described.

For sufficiently large $|\omega|$, all the pole terms
[terms in the sum of Eq.~(\ref{eq:CauchyDispersion})]
are small unless $\omega$ is near one of the poles---otherwise the terms cannot sum to one (to make $D=0$).
In the limit of large $|\omega|$, we expect solutions for
$\omega$ to be ever closer to the poles.  While this leads to 
frequencies with arbitrarily
large real parts, these unphysical modes are 
strongly damped (because Im$[\omega]\ll 0$ at these poles) and, 
as far as we know, do not affect simulation.

For a given $k\Delta x$, $v_B/\omega_p \Delta x$, and $\lambda_D/\Delta x$,
we searched for all solutions $\omega$ within
$-10< \textrm{Re}[\omega/\omega_p] < 10$, and
$-11 \pi \lambda_D/\Delta x < \textrm{Im}[\omega/\omega_p] < 20$.
Here the lower imaginary bound ensures that the region includes 
about 10 poles, or about 5 with $\textrm{Re}[\omega]>0$ 
(the vertical or imaginary spacing between poles is
$\Delta \textrm{Im}[\omega]/\omega_p = 2\pi \lambda_D/\Delta x$).
Extending the bounds just yields more solutions that are very close
to poles, hence strongly damped;
moreover, despite searching over much larger bounds in a number
of cases, we never found a mode with growth rate above
Im$[\omega]>\omega_p$.
Thus this region includes all the modes of interest---all physical modes
(corresponding to $g=0$) and all unphysical growing modes.

Using Mathematica, 
one core of a desktop computer calculates a solution in a time ranging from
a fraction of a second for large 
$v_B/\omega_p \Delta x$ to about ten minutes or longer
for some cases with small $v_B/\omega_p \Delta x$.
The results are shown in~\S\ref{sec:dispersionResults}.

%\bibliographystyle{authorBoldYearTitleDoiArxivAdsInYear}
%\bibliography{smoothingGridHeating.bib}

\newcommand{\SortNoop}[1]{}

\end{document}